\documentclass[11pt,a4paper]{article}
\usepackage[a4paper,margin=1in]{geometry}
\usepackage[T1]{fontenc}
\usepackage[utf8]{inputenc}
\usepackage{lmodern}
\usepackage{microtype}
\usepackage{amsmath,amsfonts,amssymb}
\usepackage{graphicx}
\usepackage{multirow}
\usepackage[numbers,sort&compress]{natbib}
\usepackage[colorlinks=true,linkcolor=blue,citecolor=blue,urlcolor=blue]{hyperref}
\usepackage{adjustbox}
\usepackage{hyperref}
\usepackage{placeins}
\newcommand{\keywords}[1]{\par\vspace{0.8em}\noindent\textbf{Keywords: }#1\par}
\title{\textbf{MoE-dqINR: A Unified Mixture-of-Experts Implicit Neural Representation Framework for Scan-Specific Dynamic and Quantitative MRI Reconstruction}}
\author{
Yinzhe Wu$^{1,2,*}$, Fanwen Wang$^{1,2}$, Zhenxuan Zhang$^{1}$, Zi Wang$^{1}$, \\
Chengyan Wang$^{5,6}$, Guang Yang$^{1,2,3,4,*}$\\[0.5em]
\small $^{1}$Department of Bioengineering and I-X, Imperial College London, London, SW7 2AZ, United Kingdom\\
\small $^{2}$Cardiovascular Research Centre, Royal Brompton Hospital, London, SW3 6NP, United Kingdom\\
\small $^{3}$National Heart and Lung Institute, Imperial College London, London, SW7 2AZ, United Kingdom\\
\small $^{4}$School of Biomedical Engineering \& Imaging Sciences, King's College London, London, WC2R 2LS,\\
\small United Kingdom\\
\small $^{5}$Shanghai Pudong Hospital and Human Phenome Institute, Fudan University, Shanghai, China\\
\small $^{6}$International Human Phenome Institute (Shanghai), Shanghai, China\\[0.3em]
\small $^{*}$Corresponding authors: \href{mailto:yinzhe.wu18@imperial.ac.uk}{yinzhe.wu18@imperial.ac.uk}; \href{mailto:g.yang@imperial.ac.uk}{g.yang@imperial.ac.uk}}
\date{}
\begin{document}
\maketitle
\begin{abstract}
Undersampled magnetic resonance imaging (MRI) reconstruction seeks to recover temporally or contrast-varying image series from incomplete multicoil k-space data while preserving state-dependent fidelity for dynamic and quantitative MRI (qMRI). Existing scan-specific implicit neural representations (INRs) often use monolithic spatiotemporal coordinate fields, explicit subspaces, motion or deformation models, calibration variables, or sequence-specific quantitative signal models. These design choices can limit flexibility in sharing spatial information while adapting image synthesis across acquisition states. Moreover, many INR-based baselines remain computationally demanding, typically requiring per-scan optimization times on the order of hundreds to thousands of seconds. We propose MoE-dqINR, a scan-specific multicoil MRI reconstruction framework that factorizes the image-domain representation into shared spatial experts and a state-conditioned routing pathway. Spatial experts encode reusable coordinate-dependent image content, whereas routing weights, conditioned on ordered acquisition states, synthesize each dynamic frame or contrast state from a common expert bank. The representation is coupled to a multicoil MRI forward model, uses the normalized state index to drive routing in both dynamic and quantitative MRI. By separating shared spatial representation from state-dependent synthesis, the framework provides an image-first architecture for dynamic and quantitative MRI while reducing scan-specific INR optimization to approximately 30 s per scan in our experiments. The proposed formulation establishes state-conditioned mixture-of-experts INR as a scan-specific multicoil MRI reconstruction prior that unifies shared spatial representation, dynamic- and qMRI-specific synthesis, and practical per-scan efficiency.
\end{abstract}
\section{Introduction} 

\begin{figure*}[t]
\centering
\includegraphics[width=\textwidth]{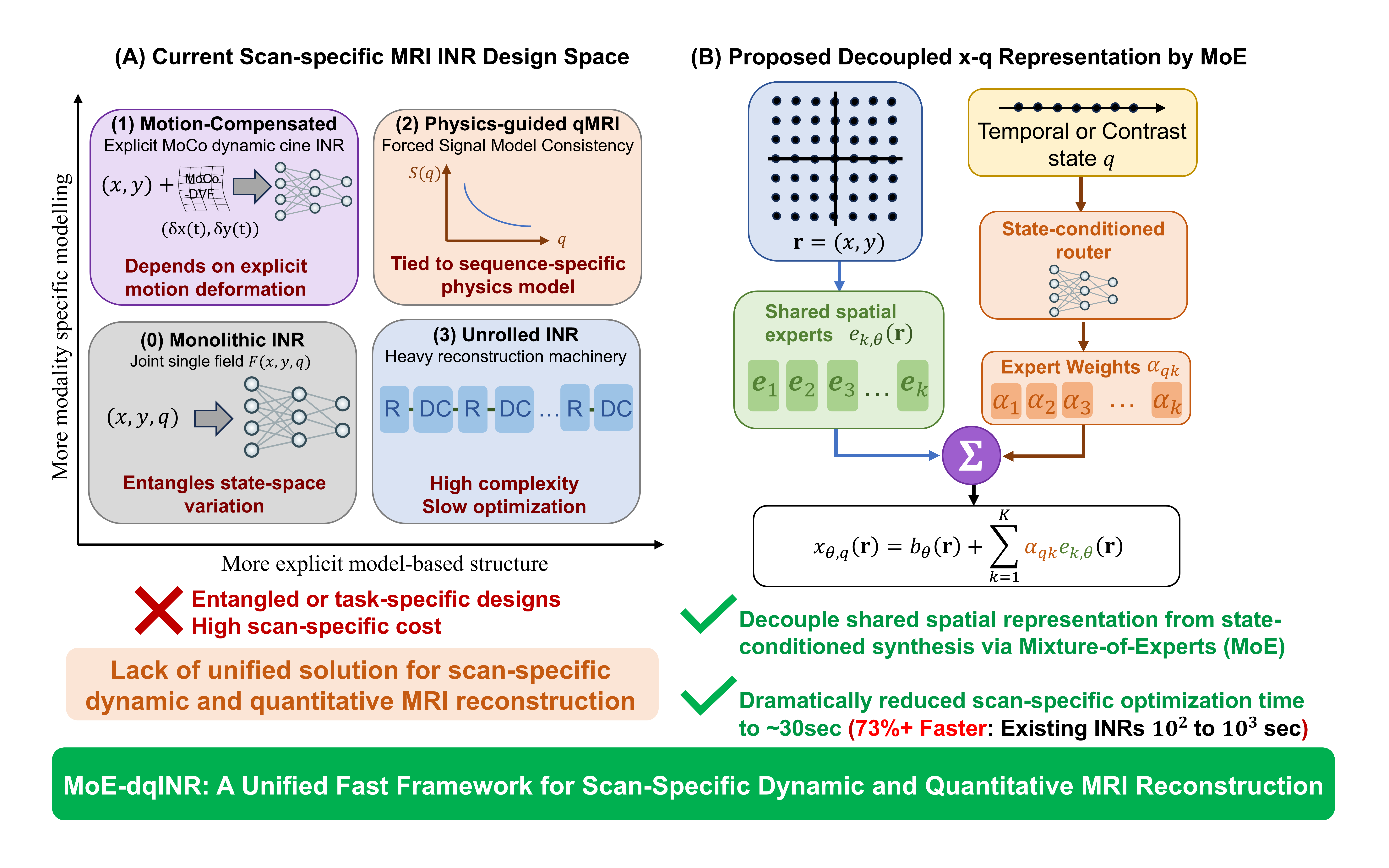}
\caption{
Overview of MoE-dqINR.
(A) Scan-specific MRI INR design space. Existing approaches either represent the full spatial--state signal with a monolithic INR, incorporate task-specific motion compensation, impose sequence-specific quantitative MRI signal models, or use unrolled reconstruction modules, leading to entangled state-space representations, limited generality, and high scan-specific optimization cost.
(B) Proposed decoupled spatial--state representation with a mixture-of-experts (MoE). Shared spatial experts $e_{k,\theta}(\mathbf{r})$ encode reusable spatial basis functions, while a $q/t$ state-conditioned router predicts expert weights $\alpha_{qk}$. The reconstructed signal is synthesized as
$x_{\theta,q}(\mathbf{r}) = b_{\theta}(\mathbf{r}) + \sum_{k=1}^{K} \alpha_{qk} e_{k,\theta}(\mathbf{r})$,
decoupling shared spatial representation from state-conditioned synthesis and enabling fast scan-specific dynamic and quantitative MRI reconstruction.
}
\label{fig:moe-dqinr-overview}
\end{figure*}
Magnetic resonance imaging reconstruction from undersampled k-space remains a central problem in both dynamic imaging and quantitative MRI. In dynamic acquisitions, the inverse problem must recover rapidly evolving spatial structure from highly incomplete measurements while preserving temporal fidelity. In quantitative MRI (qMRI), the challenge is compounded by the need to preserve contrast evolution across inversion time (TI), or other encoding axes so that downstream parameter estimation remains reliable.

Classical model-based dynamic MRI reconstruction methods have shown that this inverse problem admits substantial exploitable structure \citep{feng_grasp-pro_2020,feng_xd-grasp_2016,feng_golden-angle_2014,lingala_accelerated_2011,otazo_low-rank_2015}. In particular, temporal redundancy \citep{feng_grasp-pro_2020,feng_xd-grasp_2016,lingala_accelerated_2011,otazo_low-rank_2015}, low-rankness \citep{lingala_accelerated_2011,otazo_low-rank_2015}, sparsity \citep{feng_grasp-pro_2020,feng_xd-grasp_2016,lingala_accelerated_2011,otazo_low-rank_2015}, and motion binning \citep{feng_xd-grasp_2016} can be leveraged to stabilize recovery from highly undersampled data. Representative examples include k-t SLR \citep{lingala_accelerated_2011}, L+S decomposition \citep{otazo_low-rank_2015}, GRASP \citep{feng_golden-angle_2014}, XD-GRASP \citep{feng_xd-grasp_2016}, and GRASP-Pro \citep{feng_grasp-pro_2020}, which remain important reference points as they expose the role of temporal organization and motion-resolved modeling in accelerated MRI.

Deep learning broadened this design space substantially. Supervised end-to-end and unrolled methods have improved reconstruction quality under matched acquisition settings, but their dependence on representative training data raises generalization concerns across anatomy, contrast, sampling pattern, and acceleration rate \citep{cukur_tutorial_2025,huang_data-_2025}. Model-based deep reconstruction \citep{aggarwal_modl_2019,schlemper2017deep,ke2021learned,wang2025deep} addresses a related limitation of unconstrained end-to-end learning by embedding learned regularization within an explicit data-consistency framework, although it does not by itself remove the need for representative training data \cite{wang2025one}. Self-supervised and scan-specific methods further reduced reliance on fully sampled references. SSDU \citep{yaman_self-supervised_2020} trains physics-guided reconstruction networks without fully sampled targets by splitting acquired k-space into data-consistency and loss subsets. ZS-SSL \citep{yaman_zero-shot_2022} extends this idea to zero-shot, subject-specific reconstruction by optimizing on a single undersampled scan using disjoint data-consistency, loss, and validation subsets. DDSS \citep{zhou_dual-domain_2022} further develops acquisition-aware self-supervision for non-Cartesian MRI by combining image-domain and k-space-domain constraints. These developments establish self-supervised reconstruction and, in the zero-shot case, scan-specific optimization as viable paradigms, while leaving open the central design question of how the dynamic or quantitative image series should be parameterized and regularized. TDDIP \citep{yoo_time-dependent_2021} represents an important related step in this direction, adapting the deep image prior paradigm to dynamic MRI by optimizing a scan-specific network directly from the target acquisition.

Scan-specific implicit neural representations (INR) provide a more explicit way to parameterize the reconstructed image series. Rather than optimizing each frame or contrast state as an independent voxel grid, coordinate-based INR methods represent the image series as a continuous function of spatial coordinates and a temporal, contrast, or quantitative encoding variable. ST-INR \citep{feng_spatiotemporal_2025} showed that a joint coordinate-based representation, combined with explicit low-rank and sparse regularization, can recover dynamic MRI from highly undersampled measurements. FMLP \citep{kunz_implicit_2024} further demonstrated that expressive coordinate encodings can improve scan-specific implicit reconstruction, while INMR \citep{feng_zero-shot_2026} uses learnable low-dimensional manifold variables to condition coordinate-based spatial representations and compactly describe dynamic temporal states. Collectively, these studies establish scan-specific coordinate-based reconstruction as a viable alternative to both classical compressed sensing and pretrained deep reconstruction.

More recent work has moved beyond monolithic joint spatiotemporal coordinate fields to better structure the representation. DA-INR \citep{baik_dynamic-aware_2026} introduces a dynamic-aware representation that organizes temporal redundancy through a canonical-space/deformation formulation, while Subspace-INR \citep{huang_subspace_2026} learns separate spatial and temporal bases and explicitly reconnects scan-specific INR to classical low-rank subspace modeling. Calibration-aware scan-specific INR has also emerged: IMJENSE \citep{feng_imjense_2024} jointly models the image and coil sensitivities in parallel MRI, and IMJ-PLUS \citep{shen_free-breathing_2026} extends this idea to free-breathing dynamic MRI reconstruction with time-dependent coil sensitivity estimation. Collectively, these methods show that scan-specific MRI reconstruction is already diversified across monolithic coordinate fields, subspace-style factorizations, and joint image/calibration modeling, rather than constituting a single undifferentiated “dynamic INR” family.

An analogous diversification is now evident on the qMRI side. Joint MAPLE \citep{heydari_joint_2024} combines scan-specific self-supervised reconstruction with a joint signal-model-consistent mapping framework. PhysINR \citep{liu_physics-guided_2026} embeds the T1-rho relaxation model directly into a scan-specific INR. SUMMIT \citep{lao_coordinate-based_2025} performs zero-shot 3D multiparametric qMRI with coordinate-based representation learning, and $\pi$MRF \citep{gong_rapid_2026} integrates global spatiotemporal neural representation with explicit quantitative physics for highly accelerated MRF. In a related but less explicitly parameter-map-centered direction, Niessen et al. \citep{niessen_inr_2026} show that scan-specific INR can also be used to jointly reconstruct complementary contrast states from shared anatomical information. Hybrid INR reconstruction methods add a further neighboring design direction. UnrollINR \citep{xu_self-supervised_2025} embeds INR regularization within a physics-guided unrolled reconstruction architecture, while DiffINR \citep{chu_highly_2025} uses implicit neural representation to guide posterior sampling of diffusion models for highly accelerated MRI reconstruction. Together, these works show that scan-specific and INR-based MRI reconstruction is increasingly being structured around quantitative signal physics \citep{gong_rapid_2026,heydari_joint_2024,lao_coordinate-based_2025,liu_physics-guided_2026}, contrast coupling \citep{niessen_inr_2026}, spatiotemporal representation \citep{gong_rapid_2026}, hybrid unrolled optimization \citep{xu_self-supervised_2025}, or generative posterior sampling \citep{chu_highly_2025}, rather than relying solely on handcrafted model-based pipelines or monolithic coordinate fields. This diversification motivates architectures that can share spatial content while allowing different acquisition states to select or combine specialized components.

At the same time, the broader neural-field literature has shown that mixtures of experts (MoE), conditional computation, and routing are established mechanisms for improving specialization and efficiency in implicit representations. Neural Implicit Dictionary \citep{wang_neural_2022} represents INRs through functional combinations of dictionary elements learned with sparse MoE-style training. Levels-of-Experts \citep{hao_implicit_2022} introduces hierarchical position-dependent weights for coordinate-based representations. MoEC \citep{zhao_moec_2023} learns a gating-based partition of the domain for implicit neural compression. Neural Experts \citep{ben-shabat_neural_2024} formulates mixture-of-experts INRs for piecewise-continuous local function learning and reports improved speed, accuracy, and memory usage over single-network INRs. KiloNeRF \citep{reiser_kilonerf_2021} provides a related decomposition-based neural-field acceleration precedent. Together, these studies show that expert routing, domain partitioning, and decomposition can improve the specialization or efficiency of implicit neural representations. These works naturally raise a key question for MRI: how should expertized neural fields be formulated for scan-specific multicoil reconstruction when the non-spatial axis corresponds to temporal or quantitative state, rather than spatial partitioning or scene decomposition?

Existing scan-specific MRI methods already explore a wide range of structured representations, including learned manifold coordinates \citep{feng_zero-shot_2026}, dynamic-aware deformation models \citep{baik_dynamic-aware_2026,nurdinova_neural_2026,tian_unsupervised_2026}, explicit spatiotemporal subspaces \citep{huang_subspace_2026}, joint image and coil sensitivity estimation \citep{feng_imjense_2024,shen_free-breathing_2026}, and signal-model-aware qMRI formulations \citep{gong_rapid_2026,heydari_joint_2024,lao_coordinate-based_2025,liu_physics-guided_2026}. However, these approaches remain fragmented across task-specific inductive biases: joint coordinate fields \citep{feng_spatiotemporal_2025,kunz_implicit_2024}, subspace bases \citep{huang_subspace_2026}, calibration variables \citep{feng_imjense_2024,shen_free-breathing_2026}, motion fields \citep{baik_dynamic-aware_2026,nurdinova_neural_2026,tian_unsupervised_2026}, and explicit sequence-specific quantitative models \citep{heydari_joint_2024,lao_coordinate-based_2025,liu_physics-guided_2026}. As a result, the core limitation is no longer the absence of structure along time- or quantitative-encoding axes, but the absence of a unified scan-specific multicoil architecture that separates reusable spatial content from state-dependent synthesis. To our knowledge, no scan-specific multicoil INR architecture has yet established an image-domain factorization that decouples shared spatial structure from state-dependent variation across both dynamic MRI and qMRI (Figure~\ref{fig:moe-dqinr-overview}).

Beyond representational differences, scan-specific INR methods also face a practical efficiency bottleneck. Unlike pretrained reconstruction networks, scan-specific INRs must be optimized for each target acquisition, so per-scan optimization time directly affects clinical and research usability. This issue has been noted in prior dynamic-MRI INR work, where long optimization time (e.g., hundreds to thousands of seconds of per-scan optimization) and hyperparameter sensitivity are identified as recurring limitations of INR-based reconstruction \citep{baik_dynamic-aware_2026,kunz_implicit_2024}. This suggests that the desired architecture should not only structure state-dependent image variation, but also reduce the optimization burden of scan-specific INR (Figure~\ref{fig:moe-dqinr-overview}).

This paper addresses the lack of a unified factorized representation and the high per-scan optimization burden by formulating scan-specific multicoil MRI reconstruction as a mixture-of-experts implicit neural representation (MoE-dqINR) with decoupled spatial representation and state-conditioned routing. Shared spatial experts encode reusable spatial image content over coordinates, while a state-conditioned routing pathway generates state-dependent mixture weights that synthesize each temporal frame or quantitative contrast state from a common expert bank. This factorization separates shared spatial content from state-dependent variation, rather than requiring a single monolithic INR to jointly absorb both spatial complexity and temporal/quantitative state dynamics within one parameterization. For dynamic MRI, the framework can be augmented with a residual branch to capture frame-specific deviations; for qMRI, the formulation does not require the network to output quantitative parameter maps as its primary representation. This distinguishes the proposed method from prior scan-specific INR formulations that primarily organize the reconstruction around joint coordinates, deformation or motion fields, explicit subspaces, calibration variables, or sequence-specific quantitative signal models. More broadly, it introduces a structured, image-first scan-specific multicoil INR that explicitly disentangles shared spatial representation from state-conditioned synthesis, providing a unified representational framework for both dynamic MRI and qMRI (instantiated as cine and T1-mapping, respectively, in this study).

The main contributions of this work are as follows:
\begin{itemize}
    \item We formulate scan-specific multicoil MRI reconstruction as a mixture-of-experts implicit neural representation with decoupled spatial representation and time/quantitative-encoding-conditioned routing, providing a common coordinate-based backbone for both dynamic MRI and qMRI.
    \item We introduce an adaptive shared-expert representation in which reusable spatial experts capture common image content and state-conditioned routing synthesizes frame- or contrast-specific images from a common expert bank.
    \item We demonstrate a favorable efficiency--performance tradeoff: MoE-dqINR is the fastest to optimize among the evaluated learned baselines, reducing per-scan optimization to approximately 30 s while maintaining superior reconstruction quality.
\end{itemize}

\section{Related Work}

\subsection{Structured Dynamic MRI Reconstruction and Scan-Specific Learning}

The methodological foundations of the present work lie in two mature lines of research: structured dynamic MRI reconstruction and scan-specific learning. k-t SLR \citep{lingala_accelerated_2011} and L+S \citep{otazo_low-rank_2015} demonstrated that dynamic image series can be stabilized through low-rank and sparse decompositions, while GRASP \citep{feng_golden-angle_2014}, XD-GRASP \citep{feng_xd-grasp_2016}, and GRASP-Pro \citep{feng_grasp-pro_2020} showed how compressed sensing, radial sampling, motion-state organization, and self-calibrating subspace structure can be combined effectively in dynamic and contrast-enhanced MRI. These methods remain directly relevant because they established the central principle that temporal, contrast-phase, and motion-state organization can be exploited explicitly rather than treated as incidental nuisance variation. The proposed method follows this principle, but adopts a different representational form: rather than optimizing explicit image-series matrices, sparse components, or handcrafted temporal models, it represents state-dependent variation through routing over shared spatial experts.

Deep reconstruction and scan-specific self-supervision extend this structured viewpoint into learned settings. MoDL \citep{aggarwal_modl_2019} introduced an unrolled data-consistent reconstruction architecture with learned regularization. SSDU \citep{yaman_self-supervised_2020} removed the need for fully sampled references by partitioning acquired k-space for self-supervised training, and ZS-SSL \citep{yaman_zero-shot_2022} further specialized this strategy to scan-specific optimization on the target acquisition itself. DDSS \citep{zhou_dual-domain_2022} pushed self-supervision toward more acquisition-aware formulations through dual-domain self-supervised reconstruction for non-Cartesian MRI. Together, these methods demonstrate that scan-specific learning can be achieved through learned regularization, measurement partitioning, acquisition-aware self-supervision, or target-scan optimization. However, they do not by themselves define how a dynamic or quantitative image series should be parameterized, which motivates the subsequent move from scan-specific learning strategies to scan-specific implicit representations.

\subsection{Scan-Specific INR for Dynamic MRI}

Scan-specific neural reconstruction for dynamic MRI includes both generator-prior and coordinate-based INR formulations. TDDIP \citep{yoo_time-dependent_2021} adapted the deep image prior paradigm to dynamic MRI by optimizing a scan-specific network from the target acquisition itself, using a low-dimensional temporal latent representation rather than an externally trained model. Coordinate-based INR methods subsequently provided a more explicit continuous parameterization of the image series. ST-INR \citep{feng_spatiotemporal_2025} represents dynamic MRI as a joint spatiotemporal coordinate field regularized by low-rank and sparse priors; FMLP \citep{kunz_implicit_2024} showed that Fourier-feature coordinate encodings improve free-breathing cardiac MRI reconstruction; and INMR \citep{feng_zero-shot_2026} uses learnable low-dimensional manifold variables to condition coordinate-based spatial representations and compactly describe dynamic temporal states. Together, these works show that scan-specific neural representations can serve as powerful reconstruction priors without external training data, while also exposing a recurring design tension between expressive joint coordinate fields and more explicitly organized temporal or state-dependent representations.

Recent work addresses that tension directly. DA-INR \citep{baik_dynamic-aware_2026} organizes temporal redundancy through a dynamic-aware canonical-space formulation, whereas Subspace-INR \citep{huang_subspace_2026} learns separate spatial and temporal bases and is therefore close to classical low-rank factorization. These methods are direct neighbors of the present work because they show that recent scan-specific INR research increasingly structures the non-spatial axis rather than relying only on a plain joint \(f(x,y,t)\) coordinate field.

A second limitation of this family is computational. Because the representation is fitted directly to each undersampled acquisition, scan-specific INR reconstruction can require substantial per-scan optimization time. Prior dynamic-MRI INR studies have explicitly identified long optimization time as a practical bottleneck, and Fourier-feature INR reconstruction has also been reported to incur high computational cost \citep{baik_dynamic-aware_2026,kunz_implicit_2024}. These observations motivate representations that retain the flexibility of scan-specific neural fields while reducing the burden of fitting a high-capacity monolithic coordinate network for every scan.

\subsection{Calibration-Aware and Motion-Aware Scan-Specific Reconstruction}

A closely related branch incorporates calibration variables or motion structure into scan-specific reconstruction. IMJENSE \citep{feng_imjense_2024} performs joint image and coil-sensitivity estimation within a scan-specific implicit representation and is naturally connected to earlier joint sensitivity estimation and autocalibration methods such as JSENSE \citep{ying_joint_2007}, regularized nonlinear inversion \citep{uecker_image_2008}, and ESPIRiT \citep{uecker_espiriteigenvalue_2014}. IMJ-PLUS \citep{shen_free-breathing_2026} extends scan-specific INR to free-breathing dynamic MRI by modeling time-dependent coil sensitivities together with dynamic image structure. MoCo-INR \citep{tian_unsupervised_2026} instead explains dynamic variation through an explicit motion-compensated decomposition. Nurdinova et al. \citep{nurdinova_neural_2026} provide a related neural space-time formulation for motion-corrected MR reconstruction, further illustrating the use of neural representations to model motion-dependent image variation. Together, these methods show that scan-specific MRI reconstruction can be structured around calibration variables, motion fields, or both, in addition to the image representation itself.

\subsection{qMRI, Multi-Contrast, and Hybrid INR Formulations}

qMRI reconstruction imposes distinct downstream fidelity requirements because the reconstructed series must preserve contrast evolution for reliable parameter estimation. Joint MAPLE \citep{heydari_joint_2024} combines scan-specific self-supervision with a joint signal-model-consistent mapping framework. PhysINR \citep{liu_physics-guided_2026}, SUMMIT \citep{lao_coordinate-based_2025}, and $\pi$MRF \citep{gong_rapid_2026} each couple scan-specific implicit representation more tightly to explicit quantitative MRI physics, though at different levels of modeling and with different target applications. Niessen et al. \citep{niessen_inr_2026} show that scan-specific INR can also exploit shared anatomical information across complementary contrast states without collapsing the problem entirely into a voxel-wise parameter model. These works establish that structured qMRI and multi-contrast scan-specific reconstruction are active methodological families rather than secondary extensions of dynamic MRI reconstruction. Hybrid INR architectures provide another neighboring class. UnrollINR \citep{xu_self-supervised_2025} embeds INR regularization within a physics-guided unrolled architecture, while DiffINR \citep{chu_highly_2025} uses an INR-guided diffusion posterior-sampling framework that incorporates the MRI physical model and data fidelity. These methods differ in architectural commitment: some use INR as an auxiliary regularizer within a larger iterative solver, while others place the quantitative signal model at the center of the reconstruction. This distinction motivates a final positioning of the proposed method relative to both MRI-specific scan-specific reconstruction and the broader literature on expertized neural fields.

For qMRI and multi-contrast reconstruction, this efficiency issue is particularly relevant because multiple contrast states must be reconstructed jointly while preserving quantitative signal evolution. Therefore, a practical scan-specific qMRI INR should reduce per-scan optimization cost without replacing the contrast-state image series by an overly restrictive parameter-map-only representation.

\subsection{Expertized Neural Fields and Conditional Computation}

The broader neural-field literature provides important non-MRI precedents for the proposed architecture. Mixture-of-experts and conditional-computation models use routing mechanisms to allocate inputs to specialized components, thereby increasing representational capacity without requiring a single monolithic network to model the entire signal. In implicit neural representations, this idea has appeared in several forms. Neural Implicit Dictionary \citep{wang_neural_2022} represents target INRs through functional combinations of learned dictionary elements trained with sparse MoE-style mechanisms. Levels-of-Experts \citep{hao_implicit_2022} introduces hierarchical position-dependent weights for coordinate-based representations. MoEC \citep{zhao_moec_2023} uses a gating network to learn domain partitions for implicit neural compression, and Neural Experts \citep{ben-shabat_neural_2024} formulates MoE-INRs for local piecewise-continuous signal representation, reporting improvements in accuracy, speed, and memory efficiency over single-network INRs. KiloNeRF \citep{reiser_kilonerf_2021} provides a related decomposition-based neural-field acceleration precedent.

These works establish expert routing, domain partitioning, and local specialization as a general mechanism for decomposing implicit representations. However, existing expertized neural fields primarily route over spatial locations, scene regions, or generic signal domains, whereas scan-specific multicoil MRI reconstruction requires routing to be coupled to undersampled k-space data, coil encoding, and non-spatial acquisition states such as temporal frame, motion state, inversion time, echo time, or quantitative contrast state. This work introduces an MRI-specific expertized neural-field formulation in which shared spatial experts encode reusable spatial image content and a state-conditioned routing pathway synthesizes state-specific temporal frames or quantitative contrast states from a common expert bank. The key distinction is therefore mechanistic: the proposed method uses state-conditioned routing over shared spatial experts as the primary scan-specific image representation, rather than organizing the reconstruction primarily through a single joint coordinate field, deformation model, explicit subspace, calibration variable, quantitative signal model, or unrolled solver.

\section{Method}\label{sec:method}

\begin{figure*}[t]
\centering
\includegraphics[width=\textwidth]{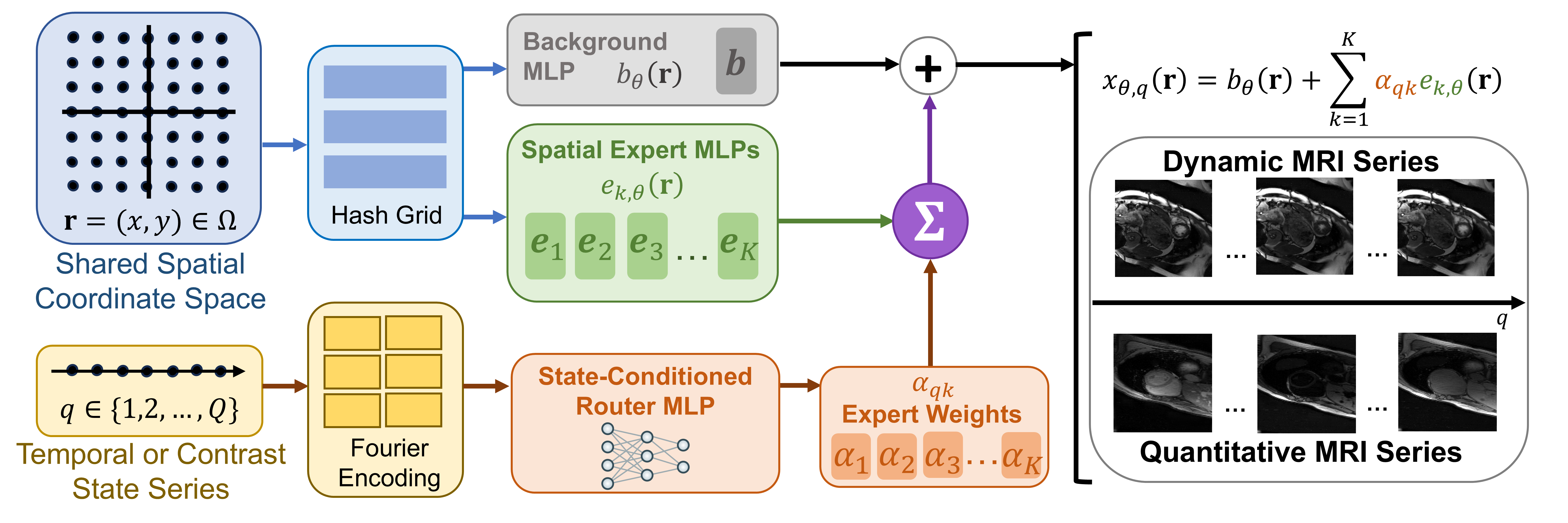}
\caption{
Main MoE-dqINR representation for dynamic and quantitative MRI reconstruction. Spatial coordinates $\mathbf r=(x,y)\in\Omega$ are shared across all acquisition states and are encoded with a hash grid to drive a background branch $b_{\theta}(\mathbf r)$ and a bank of reusable spatial expert branches $e_{k,\theta}(\mathbf r)$. The non-spatial state index $q$ (i.e., temporal frame in dynamic MRI or contrast/acquisition state in qMRI) is separately Fourier encoded and passed to a state-conditioned router that predicts expert mixture weights $\alpha_{qk}$. For each state, the reconstructed image is synthesized as
$x_{\theta,q}(\mathbf r)=b_{\theta}(\mathbf r)+\sum_{k=1}^{K}\alpha_{qk}e_{k,\theta}(\mathbf r)$,
thereby decoupling shared spatial representation from state-dependent image synthesis and enabling unified reconstruction across dynamic and quantitative MRI series.
}
\label{fig:moe-dqinr-method}
\end{figure*}

Figure~\ref{fig:moe-dqinr-method} summarizes the main MoE-dqINR reconstruction pipeline used throughout this section.

\subsection{Problem Formulation}\label{subsec:problem-formulation}

Let $q\in\{1,\ldots,Q\}$ index the non-spatial acquisition state, and let $\tau_q$ denote the acquisition value associated with state $q$. For temporal dynamic MRI, $q$ indexes cardiac temporal frames and $\tau_q$ denotes the corresponding time point. For quantitative MRI (qMRI), $q$ indexes the acquired contrast states, ordered according to the relevant acquisition parameter, and $\tau_q$ denotes that parameter value associated with state $q$ (e.g., inversion time (TI) for T1-mapping, spin-lock time (TSL) for T1-rho, or other sequence-dependent encoding variable). The state-conditioning pathway uses the normalized state index $\widetilde{q}=(q-1)/(Q-1)$ rather than $\tau_q$ directly. For each state, the unknown complex-valued image is denoted by $x_q:\Omega\rightarrow\mathbb{C}$, where $\mathbf r=(x,y)\in\Omega\subset\mathbb{R}^2$ is a spatial coordinate on the reconstructed slice. Its discretized image vector is denoted by $\mathbf x_q\in\mathbb{C}^{P}$, where $P=H\times W$ is the number of spatial samples on the reconstruction grid.

For a multicoil acquisition with $N_C$ receiver coils, the undersampled k-space measurement for coil $c$ and state $q$ is modeled as
\begin{equation}\label{eq:multicoil-forward-model}
\mathbf y_{q,c}
=
\mathbf M_q \mathbf F \mathbf S_c \mathbf x_q
+
\mathbf n_{q,c},
\qquad
c=1,\ldots,N_C .
\end{equation}
where $\mathbf S_c$ is the diagonal coil-sensitivity operator for coil $c$, $\mathbf F$ is the Fourier transform, $\mathbf M_q$ is the state-dependent sampling operator, and $\mathbf n_{q,c}$ denotes measurement noise. After stacking the measurements from all coils at state $q$, Eq.~\eqref{eq:multicoil-forward-model} can be written compactly as
\begin{equation}\label{eq:stacked-multicoil-forward-model}
\mathbf y_q
=
\mathbf E_q \mathbf x_q
+
\mathbf n_q,
\qquad
\mathbf E_q
=
\begin{bmatrix}
\mathbf M_q\mathbf F\mathbf S_1\\
\vdots\\
\mathbf M_q\mathbf F\mathbf S_{N_C}
\end{bmatrix}.
\end{equation}

The reconstruction target is the state-indexed image series
\begin{equation}\label{eq:state-indexed-image-series}
\mathbf X
=
\begin{bmatrix}
\mathbf x_1^\top\\
\vdots\\
\mathbf x_Q^\top
\end{bmatrix}
\in \mathbb C^{Q\times P},
\end{equation}
which must be estimated from the undersampled multicoil measurements $\{\mathbf y_q\}_{q=1}^{Q}$. Since each state is only partially sampled, the inverse problem is ill posed without additional structure. A conventional regularized reconstruction can be written as
\begin{equation}\label{eq:regularized-reconstruction-objective}
\widehat{\mathbf X}
=
\arg\min_{\mathbf X}
\frac{1}{2}
\sum_{q=1}^{Q}
\left\|
\mathbf E_q\mathbf x_q-\mathbf y_q
\right\|_2^2
+
\lambda \mathcal R(\mathbf X),
\end{equation}
where the first term enforces multicoil k-space data consistency and $\mathcal R(\mathbf X)$ denotes prior information on the image series. Classical dynamic and qMRI methods instantiate $\mathcal R(\mathbf X)$ using temporal sparsity, low-rank structure, motion consistency, subspace constraints, or signal-evolution models.

In the proposed method, the primary prior is instead imposed through the image parameterization itself. Specifically, we represent the image series using a scan-specific mixture-of-experts implicit neural representation (MoE-dqINR), rather than optimizing each $\mathbf x_q$ as an independent image vector or representing the series by a monolithic coordinate field $f_\theta(\mathbf r,\widetilde{q})$. Let $\theta$ denote the trainable parameters of the spatial encoder, background branch, expert branches, and state-conditioned routing network. The reconstructed image at state index $q$ is written as
\begin{equation}\label{eq:moe-inr-synthesis-function}
x_{\theta,q}(\mathbf r)
=
\mathcal G_\theta(\mathbf r,\widetilde{q}),
\end{equation}
where $\mathcal G_\theta$ denotes the proposed MoE-dqINR synthesis function and $\widetilde{q}$ is the normalized state index used for conditioning. In the core model, this synthesis function decomposes each state image into a shared spatial background and a state-conditioned mixture of shared spatial experts:
\begin{equation}\label{eq:moe-inr-core-decomposition}
\mathcal G_\theta(\mathbf r,\widetilde{q})
=
b_\theta(\mathbf r)
+
\sum_{k=1}^{K}
\alpha_k(\widetilde{q})\,
e_{k,\theta}(\mathbf r).
\end{equation}
Here, $b_\theta(\mathbf r)$ is the shared background branch, $e_{k,\theta}(\mathbf r)$ is the $k$-th spatial expert, $K$ is the number of experts, and $\alpha_k(\widetilde{q})$ is the routing weight assigned to expert $k$ at normalized state index $\widetilde{q}$. For compactness, we write $\alpha_{qk}=\alpha_k(\widetilde{q})$. The routing weights satisfy
\begin{equation}\label{eq:routing-simplex-constraints}
\alpha_{qk}\ge 0,
\qquad
\sum_{k=1}^{K}\alpha_{qk}=1.
\end{equation}

Evaluating $x_{\theta,q}(\mathbf r)$ on the reconstruction grid
$\Omega_h=\{\mathbf r_p\}_{p=1}^{P}$ gives the discretized image vector

\begin{equation}\label{eq:discretized-parameterized-image}
\begin{aligned}
\mathbf x_{\theta,q}
&=
\left[
x_{\theta,q}(\mathbf r_1),\ldots,
x_{\theta,q}(\mathbf r_P)
\right]^\top \\
&=
\left[
\mathcal G_\theta(\mathbf r_1,\widetilde{q}),\ldots,
\mathcal G_\theta(\mathbf r_P,\widetilde{q})
\right]^\top
\in\mathbb C^{P}.
\end{aligned}
\end{equation}

Substituting this parameterized image into the multicoil forward model yields the scan-specific neural reconstruction problem
\begin{equation}\label{eq:scan-specific-neural-reconstruction-objective}
\widehat{\theta}
=
\arg\min_{\theta}
\frac{1}{2}
\sum_{q=1}^{Q}
\left\|
\mathbf E_q
\mathbf x_{\theta,q}
-
\mathbf y_q
\right\|_2^2 .
\end{equation}

This formulation separates the physical acquisition model from the learned representation. The operator $\mathbf E_q$ enforces consistency with the acquired multicoil k-space data, whereas the MoE-dqINR parameterization constrains each state image to be synthesized from a shared spatial background and a state-conditioned mixture of reusable spatial experts. Dynamic and quantitative image states are therefore reconstructed by routing over a shared spatial expert bank, thereby decoupling spatial representation from state-dependent image synthesis under the multicoil forward model.

Figure~\ref{fig:moe-dqinr-overview} summarizes this decoupled architecture and its coupling to the multicoil reconstruction pipeline.

\subsection{Coordinate Encoding and MoE-dqINR Architecture}\label{subsec:coordinate-encoding-and-moe-inr-architecture}

The proposed reconstruction model parameterizes the image series through a scan-specific image-domain MoE-dqINR. The spatial coordinate $\mathbf r=(x,y)$ and the normalized state index $\widetilde{q}$ enter the model through separate pathways. Spatial coordinates are used to evaluate shared coordinate-based image components, whereas $\widetilde{q}$ is used only to determine how these components are combined for each state. This separation is central to the proposed formulation: spatial anatomy is represented by reusable expert functions, while temporal or quantitative variation is expressed by state-conditioned routing.

For a reconstructed image grid of size $H\times W$, pixel locations are first normalized to the unit square,
\begin{equation}\label{eq:normalized-pixel-coordinates}
\begin{aligned}
\mathbf r_{ij}
&=
\left(
\frac{j}{W-1},
\frac{i}{H-1}
\right),
&\qquad
i=0,\ldots,H-1,\quad
j=0,\ldots,W-1 .
\end{aligned}
\end{equation}

The normalized coordinate is then mapped to a multiscale feature vector through a learned spatial encoder,
\begin{equation}\label{eq:learned-spatial-encoder}
\boldsymbol\psi(\mathbf r)=\Phi_{\xi}(\mathbf r),
\end{equation}
where $\Phi_{\xi}$ denotes the multiresolution hash-grid encoder and $\xi$ denotes its trainable parameters. The encoded feature $\boldsymbol\psi(\mathbf r)$ is shared by the background branch and all expert branches. This design allows the scan-specific neural field to represent fine spatial detail efficiently while keeping the state variable outside the spatial encoder.

The encoded coordinate is passed to a shared background branch and a bank of $K$ spatial expert branches. Let $f_{\nu}$ denote a generic image-generating branch, where $\nu=0$ corresponds to the background branch and $\nu=k$ corresponds to the $k$-th expert branch. Each branch is implemented as a fully connected multi-layer perceptron (MLP) with $D_{\nu}$ hidden layers and SiLU activations \citep{elfwing_sigmoid-weighted_2018}. Its computation is
\begin{equation}\label{eq:branch-hidden-initialization}
\mathbf h_{\nu}^{(0)}=\boldsymbol\psi(\mathbf r),
\end{equation}

\begin{equation}\label{eq:branch-hidden-layer-update}
\begin{aligned}
\mathbf h_{\nu}^{(\ell+1)}
&=
\operatorname{SiLU}
\left(
\mathbf W_{\nu}^{(\ell)}
\mathbf h_{\nu}^{(\ell)}
+
\mathbf a_{\nu}^{(\ell)}
\right),
&\qquad
\ell=0,\ldots,D_{\nu}-1 .
\end{aligned}
\end{equation}

\begin{equation}\label{eq:branch-output-projection}
f_{\nu}(\boldsymbol\psi(\mathbf r))
=
\mathbf W_{\nu}^{(D_{\nu})}\mathbf h_{\nu}^{(D_{\nu})}
+
\mathbf a_{\nu}^{(D_{\nu})}.
\end{equation}

Here, $\mathbf h_{\nu}^{(\ell)}$ denotes the hidden feature vector at layer $\ell$, $\mathbf W_{\nu}^{(\ell)}$ and $\mathbf a_{\nu}^{(\ell)}$ denote the weight matrix and bias vector of layer $\ell$, and depth $D_{\nu}$ is the number of hidden layers in branch $f_{\nu}$, excluding the final linear output projection. The final layer outputs two real-valued channels, interpreted as the real ($\mathrm{Re}$) and imaginary ($\mathrm{Im}$) parts of a complex image value.

The background branch output is written as
\begin{equation}\label{eq:background-branch-complex-output}
b_{\theta}(\mathbf r)
=
b_{\theta,\mathrm{Re}}(\mathbf r)
+
i\,b_{\theta,\mathrm{Im}}(\mathbf r),
\end{equation}
and the $k$-th expert output is written as
\begin{equation}\label{eq:expert-branch-complex-output}
e_{k,\theta}(\mathbf r)
=
e_{k,\theta,\mathrm{Re}}(\mathbf r)
+
i\,e_{k,\theta,\mathrm{Im}}(\mathbf r),
\qquad
k=1,\ldots,K .
\end{equation}

The background branch represents common spatial content shared across all acquisition states, while the expert branches represent reusable spatial components that can be recombined differently as a function of $\widetilde{q}$. Importantly, the experts are spatial functions only: they do not directly receive $\widetilde{q}$ as an input. State dependence is introduced through the routing weights defined in the following subsection.

\subsection{State Encoding and Expert Routing}\label{subsec:state-encoding-and-expert-routing}

The separate state-conditioning pathway produces the routing weights in Eq.~\eqref{eq:routing-simplex-constraints} from the normalized state index $\widetilde{q}$ rather than from the raw acquisition value $\tau_q$. For qMRI, this means the contrast states are first ordered by acquisition parameter and then indexed by $q$. Let $\boldsymbol\varphi(\widetilde{q})$ denote the encoded state feature used by the routing network. In both dynamic and qMRI, we use a deterministic Fourier state encoding of the normalized state index. With $H_s$ sine-cosine frequency pairs, the state feature is
\begin{equation}\label{eq:fourier-state-feature}
\begin{aligned}
\boldsymbol\varphi(\widetilde{q})
=
\big[&
\widetilde{q},
\sin(2\pi\widetilde{q}),
\cos(2\pi\widetilde{q}),
\ldots, 
\sin(2\pi H_s\widetilde{q}),
\cos(2\pi H_s\widetilde{q})
\big]^{\top}.
\end{aligned}
\end{equation}

This encoding is analogous to the Fourier feature encodings commonly used in coordinate-based neural representations, but here it is applied to the one-dimensional acquisition-state index and is used only by the routing network.

For dynamic, this Fourier state encoding naturally represents periodic variation across the cardiac cycle. For qMRI, the same harmonic expansion acts as a richer positional encoding over the ordered contrast states. In the experiments reported here, $H_s=2$ for both dynamic and qMRI.

Thus, the same routing mechanism and the same Fourier state encoding are used for temporal frames and quantitative contrast states; only the acquisition-specific interpretation of the ordered state index differs.

\vspace{\baselineskip}

\textbf{The routing network} $g_{\omega}$, parameterized by $\omega$, maps the state feature to $K$ expert logits,
\begin{equation}\label{eq:routing-network-logits}
\mathbf z_q
=
g_{\omega}\!\left(\boldsymbol\varphi(\widetilde{q})\right)
\in\mathbb R^{K}.
\end{equation}

The logits used for expert selection are produced directly from $\boldsymbol\varphi(\widetilde{q})$. The routing weights are obtained by softmax normalization,
\begin{equation}\label{eq:softmax-routing-weights}
\alpha_{qk}
=
\frac{\exp(z_{qk})}
{\sum_{j=1}^{K}\exp(z_{qj})},
\qquad
k=1,\ldots,K.
\end{equation}

The expert mixture at each state is therefore determined by applying the routing network to the Fourier state encoding of the normalized state index and softmax-normalizing the resulting logits. This gives the core reconstruction model
\begin{equation}\label{eq:core-reconstruction-model}
x_{\theta,q}(\mathbf r)
=
b_{\theta}(\mathbf r)
+
\sum_{k=1}^{K}
\alpha_{qk}
e_{k,\theta}(\mathbf r).
\end{equation}

This formulation differs from a joint coordinate field $f_{\theta}(\mathbf r,\widetilde{q})$. The state variable does not modulate the spatial encoder or the expert MLPs directly; instead, it controls the mixture coefficients assigned to spatial functions that are shared across all states.

The learned state-conditioned routing behavior is analyzed in Section~\ref{subsec:routing-analysis}, where we examine whether the simplex-constrained router produces distributed temporal mixing for dynamic MRI and contrast-selective expert allocation for T1 mapping.

\subsection{Dynamic Temporal Residual Branch}\label{subsec:dynamic-temporal-residual-branch}

For dynamic reconstruction, the core MoE synthesis in Eq.~\eqref{eq:core-reconstruction-model} is augmented by a compact state-conditioned residual branch. This branch is included to increase expressive capacity for temporal dynamics, but it is not part of the core representation used for qMRI.

When enabled, the reconstructed image is
\begin{equation}\label{eq:dynamic-temporal-residual-branch}
x_{\theta,q}(\mathbf r)
=
b_{\theta}(\mathbf r)
+
\sum_{k=1}^{K}
\alpha_{qk}e_{k,\theta}(\mathbf r)
+
\gamma
\sum_{\ell=1}^{L}
\beta_{q\ell}
r_{\ell,\theta}(\mathbf r),
\end{equation}
where $r_{\ell,\theta}(\mathbf r)$ denotes the $\ell$-th residual spatial basis function, $\beta_{q\ell}$ is its state-dependent coefficient, $L$ is the residual rank, and $\gamma$ is a fixed residual scaling factor. The residual basis functions are generated from the same spatial feature $\boldsymbol\psi(\mathbf r)$ as the background and expert branches, while the residual coefficients are generated from the Fourier state feature $\boldsymbol\varphi(\widetilde{q})$. Thus, the residual branch preserves the same decoupled structure as the main model: spatial functions depend on $\mathbf r$, and state dependence enters through low-dimensional coefficients.

In the dynamic configuration, the residual branch is deliberately compact, with $L=2$, a small residual scale $\gamma=0.022$, and substantially smaller residual MLPs than the main expert branches; the full architectural settings are summarized in Section~\ref{subsec:implementation-details}. This design makes the residual path a low-capacity additive correction rather than a second full expert bank. Importantly, the residual coefficients $\beta_{q\ell}$ are not softmax-normalized. Unlike the main routing weights $\alpha_{qk}$, which are nonnegative and sum to one, the residual coefficients are free signed outputs of the state-conditioning branch. The residual path can therefore add or subtract small state-dependent corrections without forcing these corrections to compete with the main experts for probability mass.

This coefficient geometry distinguishes the residual branch from simply increasing the number of main experts. The main MoE pathway uses simplex-constrained routing to synthesize each frame from dominant shared spatial experts, whereas the residual branch provides a small unconstrained additive subspace for structured temporal residual variation. It is intended to absorb low-dimensional, low-amplitude dynamic-MRI-specific deviations that are not efficiently represented by reallocating the main expert mixture. It is not a full spatiotemporal coordinate field and does not estimate motion. Instead, it serves as a lightweight augmentation to the main MoE structure for dynamic reconstruction, increasing temporal expressivity while preserving the core principle of spatial representation through reusable experts and state-conditioned routing.

\subsection{Training Objective}\label{subsec:training-objective}

The reconstruction is trained by minimizing k-space data inconsistency under the multicoil forward model over all acquisition states $q\in\{1,\ldots,Q\}$. For a state $q\in\{1,\ldots,Q\}$, the predicted k-space data for coil $c$ are
\begin{equation}\label{eq:predicted-k-space-data}
\widehat{\mathbf y}_{q,c}
=
\mathbf M_q\mathbf F\mathbf S_c\mathbf x_{\theta,q}.
\end{equation}

The data-consistency loss is defined as the complex squared residual over the acquired k-space samples,
\begin{equation}\label{eq:data-consistency-loss}
\mathcal L_{\mathrm{dc}}(\theta)
=
\sum_{q=1}^{Q}
\sum_{c=1}^{N_C}
\left\|
\mathbf M_q\mathbf F\mathbf S_c\mathbf x_{\theta,q}
-
\mathbf y_{q,c}
\right\|_2^2 ,
\end{equation}
where the norm in Eq.~\eqref{eq:data-consistency-loss} is evaluated over the acquired k-space samples, with the complex residual measured by the squared magnitude.

In the present implementation, both dynamic and qMRI use all available acquisition states in every update, so the loss is always accumulated over the full state set $\{1,\ldots,Q\}$.

In addition to the data-consistency term, optimization applies weight decay to the image-generating parameter group. Recall that $\theta$ was introduced before Eq.~\eqref{eq:moe-inr-synthesis-function} as the collection of all trainable parameters in the reconstruction model. Here, we define $\theta_{\mathrm{img}}\subset\theta$ as the subset that directly generates spatial image content: the spatial-encoder parameters $\xi$ in Eq.~\eqref{eq:learned-spatial-encoder}, the branch parameters in Eqs.~\eqref{eq:branch-hidden-initialization}--\eqref{eq:branch-output-projection} that produce the background branch $b_{\theta}(\mathbf r)$ and expert branches $e_{k,\theta}(\mathbf r)$ in Eqs.~\eqref{eq:background-branch-complex-output} and \eqref{eq:expert-branch-complex-output}, and, for dynamic when the residual path is enabled, the parameters of the residual spatial basis functions $r_{\ell,\theta}(\mathbf r)$ in Eq.~\eqref{eq:dynamic-temporal-residual-branch}.
Let $\lambda_{\mathrm{wd}}$ denote the corresponding weight-decay coefficient. In the present implementation, $\lambda_{\mathrm{wd}}=1.2\times10^{-4}$ for dynamic reconstruction and $\lambda_{\mathrm{wd}}=10^{-4}$ for qMRI. The effective training objective is therefore
\begin{equation}\label{eq:effective-training-objective}
\mathcal L
=
\mathcal L_{\mathrm{dc}}
+
\lambda_{\mathrm{wd}}\|\theta_{\mathrm{img}}\|_2^2,
\end{equation}
where the $\ell_2$ penalty acts only on the image-generating parameter group.

\subsection{Optimization Schedule}\label{subsec:optimization-schedule}

The model is optimized separately for each scan. All trainable components are updated jointly from the first iteration, including the spatial hash-grid encoder, the background branch, the expert branches, the routing MLP $g_{\omega}$, and, for dynamic reconstruction, the residual branch described in Eq.~\eqref{eq:dynamic-temporal-residual-branch}.

Training runs for 850 iterations. The image-generating network is optimized with initial learning rate $10^{-3}$, and the routing pathway is optimized with initial learning rate $2\times10^{-3}$. As reflected in Eq.~\eqref{eq:effective-training-objective}, weight decay is applied only to the image-generating parameter group; no weight decay is applied to the routing or time-conditioning parameter groups.

\subsection{Implementation Details}\label{subsec:implementation-details}

The same decoupled MoE-dqINR formulation is used for dynamic reconstruction (here specifically instantiated as cine) and qMRI (here specifically instantiated as T1-mapping). Both modalities use the same Fourier state encoding of the normalized state index in the routing pathway and the same effective training objective in Eq.~\eqref{eq:effective-training-objective}, while model capacity and residual-branch usage differ by modality. All reported training was performed on a single NVIDIA GeForce RTX 3090 GPU. The exact architectural settings used in the reported reconstructions are summarized below to make the implementation reproducible.

\vspace{\baselineskip}

\noindent\textbf{For dynamic MRI reconstruction (instantiated as cine in this study),} $q$ indexes temporal frames. The routing feature uses the Fourier state encoding of the normalized state index in Eq.~\eqref{eq:fourier-state-feature} with $H_s=2$. The spatial encoder is a multiresolution hash-grid encoder with 18 levels, 2 features per level, $\log_2$ hash-map size 20, base resolution 20, and per-level scale 1.45. The model uses $K=7$ spatial experts. The background branch has width 160 and depth $D_{\nu}=6$, and each MLP in the expert branch has width 224 and depth $D_{\nu}=7$. The effective training objective combines multicoil k-space data consistency over all temporal frames with weight decay on the image-generating parameter group.

\vspace{\baselineskip}

\noindent\textbf{For qMRI (instantiated as T1-mapping in this study),} $q$ indexes the acquired contrast state images ordered by inversion time (TI). The routing feature uses the same Fourier state encoding of the normalized state index in Eq.~\eqref{eq:fourier-state-feature} with $H_s=2$. The spatial encoder is a multiresolution hash-grid encoder with 16 levels, 2 features per level, $\log_2$ hash-map size 19, base resolution 16, and per-level scale 1.5. The model uses $K=10$ spatial experts. The background branch has width 128 and depth $D_{\nu}=5$, and each expert branch has width 160 and depth 6. The effective training objective combines q-normalized multicoil k-space data consistency over all contrast states with weight decay on the image-generating parameter group.

\section{Experiments}\label{sec:experiments}

\subsection{Dataset}\label{subsec:dataset}

We evaluated the proposed method on retrospectively undersampled dynamic cine and T1-mapping data derived from the CMRxRecon dataset \citep{wang_cmrxrecon_2024} and OCMR dataset \citep{chen_ocmr_2020, noauthor_ocmr_nodate}.

As per the standard CMRxRecon setup \citep{wang_cmrxrecon_2024}, data were acquired on a 3\,T MAGNETOM Vida scanner (Siemens Healthineers, Germany) using a 32-channel receiver coil. Cine imaging used 2D TrueFISP readout in the short-axis view with retrospective ECG-gated segmented acquisition under breath-hold. Cine short-axis scans typically contain 12 frames, spatial resolution of $1.5\times1.5$\,mm$^2$, slice thickness of 8.0\,mm, repetition time (TR) of 3.6\,ms, and echo time (TE) of 1.6\,ms. T1 mapping was acquired in the short-axis view using a MOLLI sequence which acquired 9 images with different T1 weightings with the 4-(1)-3-(1)-2 scheme, with ECG triggering at end-diastole. The typical T1-mapping parameters are spatial resolution of $1.5\times1.5$\,mm$^2$, slice thickness of 5.0\,mm, TR of 2.7\,ms, and TE of 1.1\,ms, with The inversion time varied among subjects according to the real-time heart rate \citep{wang_cmrxrecon_2024}.

For retrospective undersampling, cine short-axis data were sampled from fully sampled cine scans over 10 subjects using VISTA Cartesian masks, whereas T1-mapping data were sampled from fully sampled cine scans over 20 subjects using Golden Angle Cartesian masks. Coil sensitivity maps were estimated with ESPIRiT from the fully sampled central k-space region using 16 auto-calibration (ACS) lines. Images were center-cropped in image space to task-specific dimensions: $204\times204$ pixels for cine short-axis and $144\times144$ pixels for T1 mapping.  For both tasks, nominal acceleration factors (AF) of $4\times$, $6\times$, and $8\times$ were evaluated (ACS not included for AF calculations).

To further assess applicability to low-field cardiac cine MRI, we additionally evaluated the model on fully sampled cine scans over 18 subjects from the OCMR dataset acquired on a 0.55\,T MAGNETOM Free.Max system (Siemens Healthineers, Erlangen, Germany) \citep{chen_ocmr_2020, noauthor_ocmr_nodate}. The OCMR Free.Max cine data were acquired using a 2D TrueFISP readout in the short-axis view with retrospective ECG-gated segmented acquisition under breath-hold. The acquisition was performed at 0.55\,T using a 10-element receiver coil array and reconstructed into 12 cardiac frames. For this additional OCMR cine evaluation, we used the same retrospective multicoil reconstruction pipeline as for CMRxRecon cine, except that nominal acceleration factors of $2\times$, $4\times$, and $6\times$ were evaluated.

\subsection{Baseline Methods}\label{subsec:Comparison-methods}

We compared MoE-dqINR with baseline methods selected to cover the main methodological families discussed in the Method and Related Work sections: classical model-based dynamic reconstruction, scan-specific coordinate-based INR, calibration-aware scan-specific reconstruction, motion-aware reconstruction, self-supervised scan-specific learning, unrolled INR-based reconstruction, and quantitative-physics-guided INR. This selection is intended to benchmark the proposed state-conditioned shared-expert factorization against methods that use different assumptions for organizing temporal or contrast-state variation.

For cine reconstruction, the comparison included ST-INR \citep{feng_spatiotemporal_2025}, GRASP-Pro \citep{feng_grasp-pro_2020}, IMJ-PLUS \citep{shen_free-breathing_2026}, IMJENSE \citep{feng_imjense_2024}, MoCo-INR \citep{tian_unsupervised_2026}, UnrollINR \citep{xu_self-supervised_2025}, ZS-SSL \citep{yaman_zero-shot_2022}, TDDIP \citep{yoo_time-dependent_2021}, and FMLP \citep{kunz_implicit_2024}. These baselines were chosen because they represent complementary strategies for accelerated dynamic MRI: GRASP-Pro provides a strong classical model-based reference with temporal organization and self-calibrating structure; ST-INR, TDDIP, and FMLP represent scan-specific neural or coordinate-based reconstruction priors; IMJENSE and IMJ-PLUS evaluate calibration-aware scan-specific formulations; MoCo-INR tests an explicit motion-compensated representation; ZS-SSL represents zero-shot self-supervised deep-learning reconstruction in an unrolled way; and UnrollINR provides a hybrid unrolled architecture with INR regularization.

For T1-mapping reconstruction, the comparison included PhysINR \citep{liu_physics-guided_2026}, GRASP-Pro \citep{feng_grasp-pro_2020}, IMJ-PLUS \citep{shen_free-breathing_2026}, IMJENSE \citep{feng_imjense_2024}, UnrollINR \citep{xu_self-supervised_2025}, ZS-SSL \citep{yaman_zero-shot_2022}, TDDIP \citep{yoo_time-dependent_2021}, and FMLP \citep{kunz_implicit_2024}. PhysINR was included as a quantitative-physics-guided INR baseline and was adapted from its original T1-rho formulation to the T1-mapping signal model used in this study. The remaining methods were included to evaluate whether the proposed image-first shared-expert routing remains competitive against representative model-based, calibration-aware, self-supervised, generator-prior, unrolled, and coordinate-based scan-specific reconstruction strategies when the non-spatial state axis corresponds to ordered T1-weighted contrast states rather than temporal frames.

\subsection{Evaluation Metrics}\label{subsec:evaluation-metrics}

For cine reconstruction, quantitative evaluation was performed using peak signal-to-noise ratio (PSNR), structural similarity index measure (SSIM), and learned perceptual image patch similarity (LPIPS) on the reconstructed cine images.

For T1-mapping reconstruction, we report per-state ($q$-wise) PSNR, SSIM, and LPIPS on the reconstructed T1-weighted images, together with normalized mean squared error (NMSE) on the derived T1 map.

\section{Results}\label{sec:results}

\subsection{Reconstruction Performance in Cine MRI and T1 Mapping}\label{subsec:reconstruction-performance}

Tables~\ref{tab:cine-main-results} and \ref{tab:t1m-main-results} summarize the main quantitative comparisons for cine MRI and T1-mapping reconstruction, respectively. For cine MRI, MoE-dqINR achieves the best PSNR, SSIM, and LPIPS across all tested acceleration factors ($4\times$, $6\times$, and $8\times$). For T1 mapping, MoE-dqINR also consistently achieves the best per-state PSNR, SSIM, and LPIPS, while producing the lowest T1-map NMSE across all tested acceleration factors. These results indicate that the shared-expert factorization remains effective under increasingly aggressive undersampling, with consistent gains in both image-domain reconstruction quality and quantitative-map fidelity.

To assess applicability beyond the main 3\,T benchmark, Table~\ref{tab:ocmr-lowfield-results} reports additional validation on the 0.55\,T OCMR cine dataset. MoE-dqINR again achieves the best PSNR, SSIM, and LPIPS at all tested acceleration factors ($2\times$, $4\times$, and $6\times$), supporting that the proposed shared-expert factorization transfers effectively to low-field cine MRI.

\begin{figure*}[!t]
\centering
\setlength{\tabcolsep}{2pt}
\renewcommand{\arraystretch}{0.75}
\begin{tabular}{c}
\textbf{(A) AF=$4\times$} \\[3mm]
\includegraphics[width=1.0\textwidth]{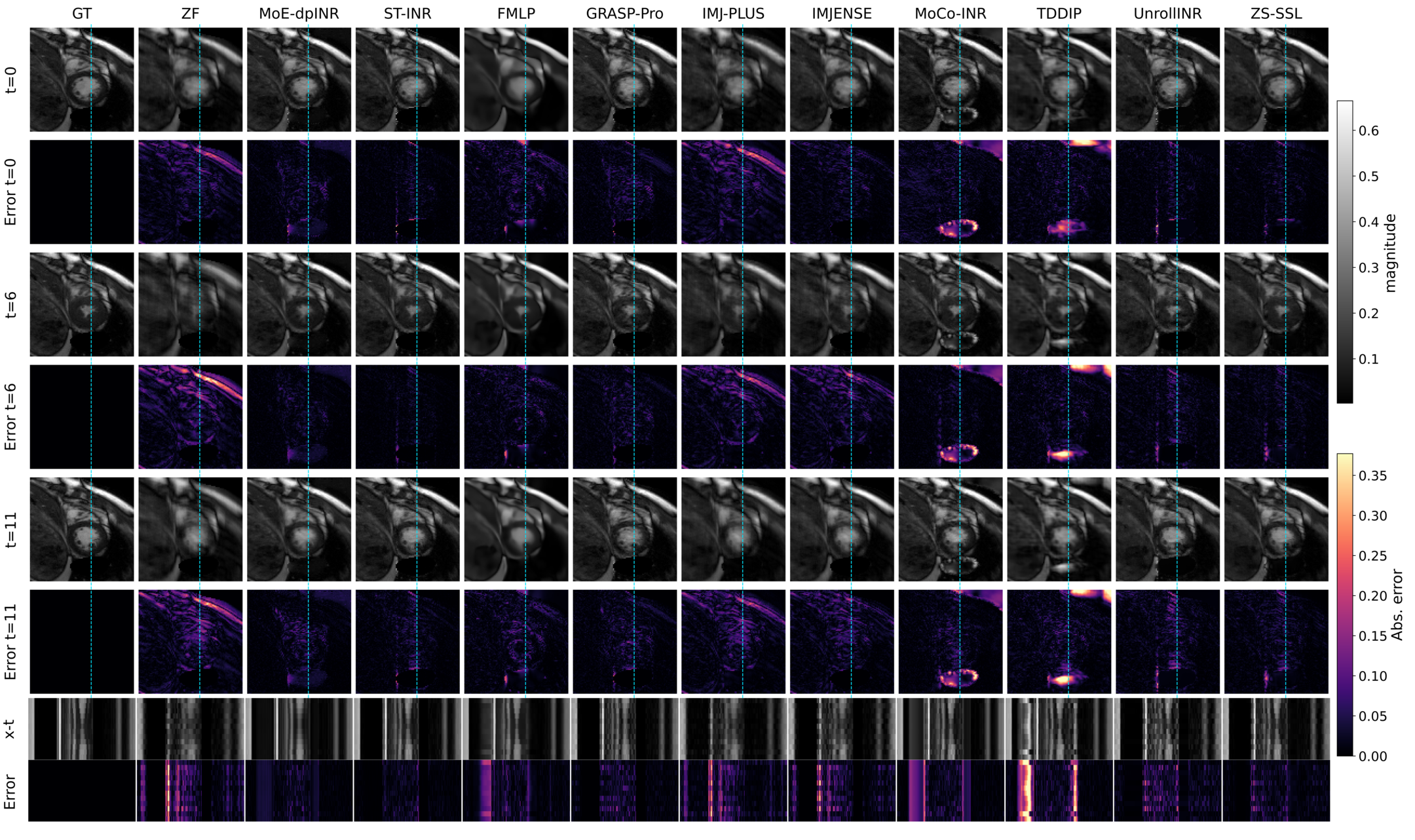} \\[1mm]
\textbf{(B) AF=$6\times$} \\[6mm]
\includegraphics[width=1.0\textwidth]{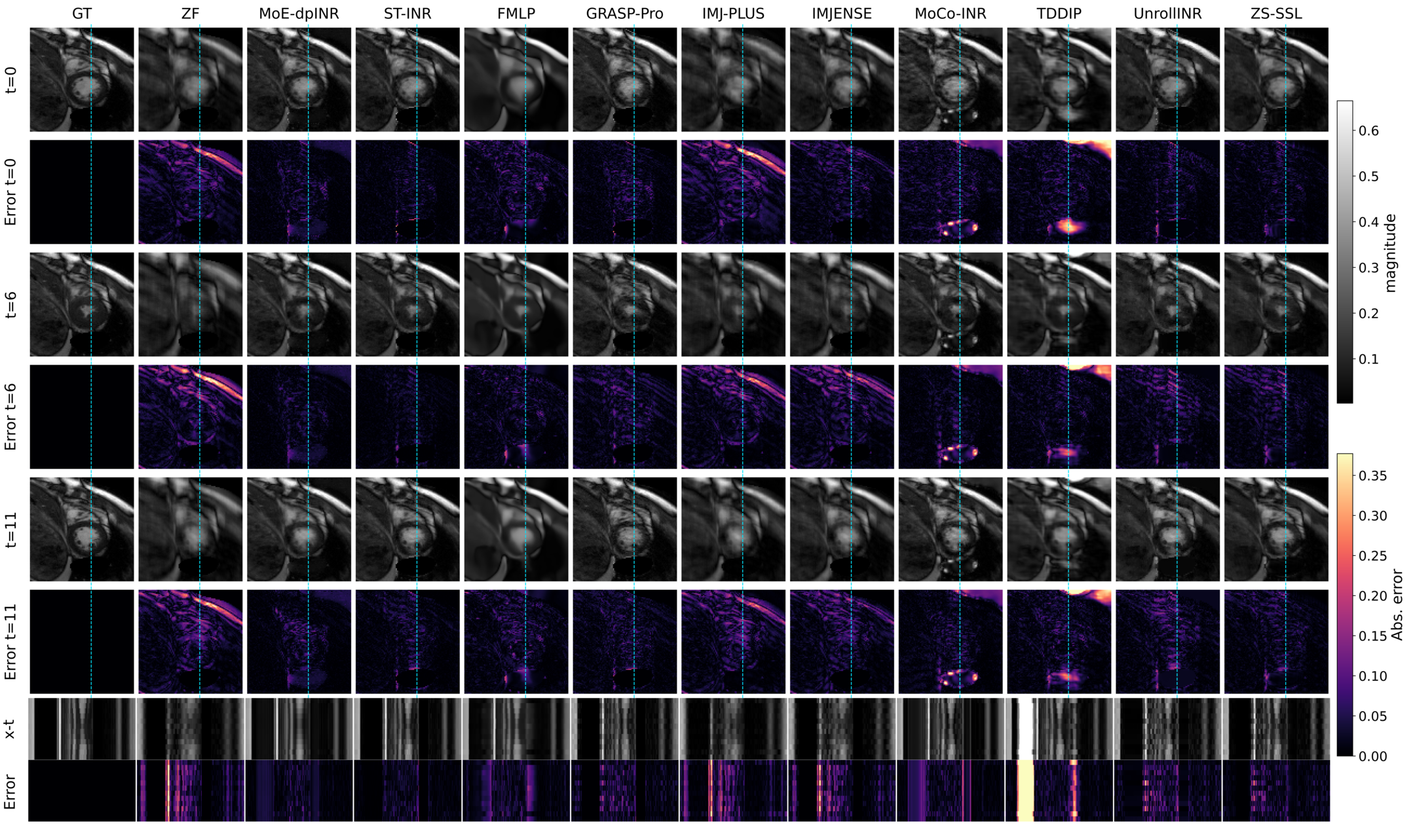}
\end{tabular}

\caption{
Representative cine reconstruction comparison (Cardiac region-of-interest (ROI)) at (A) AF=$4\times$ and (B) AF=$6\times$. Each panel shows local Cardiac ROI and corresponding $x$--$t$ cross-sections, enabling comparison of spatial detail and temporal consistency across the cardiac cycle. MoE-dqINR preserves myocardial structure and temporal continuity with lower localized error than the compared baselines.
}
\label{fig:cine-qualitative-roi}
\end{figure*}

\begin{figure*}[!t]
\centering
\setlength{\tabcolsep}{0pt}
\renewcommand{\arraystretch}{0.65}

\begin{adjustbox}{max width=\textwidth, max totalheight=0.85\textheight, keepaspectratio}
\begin{tabular}{c}
\textbf{(A) AF=$4\times$: T1-weighted image reconstruction} \\[0.5mm]
\includegraphics[width=0.98\textwidth]{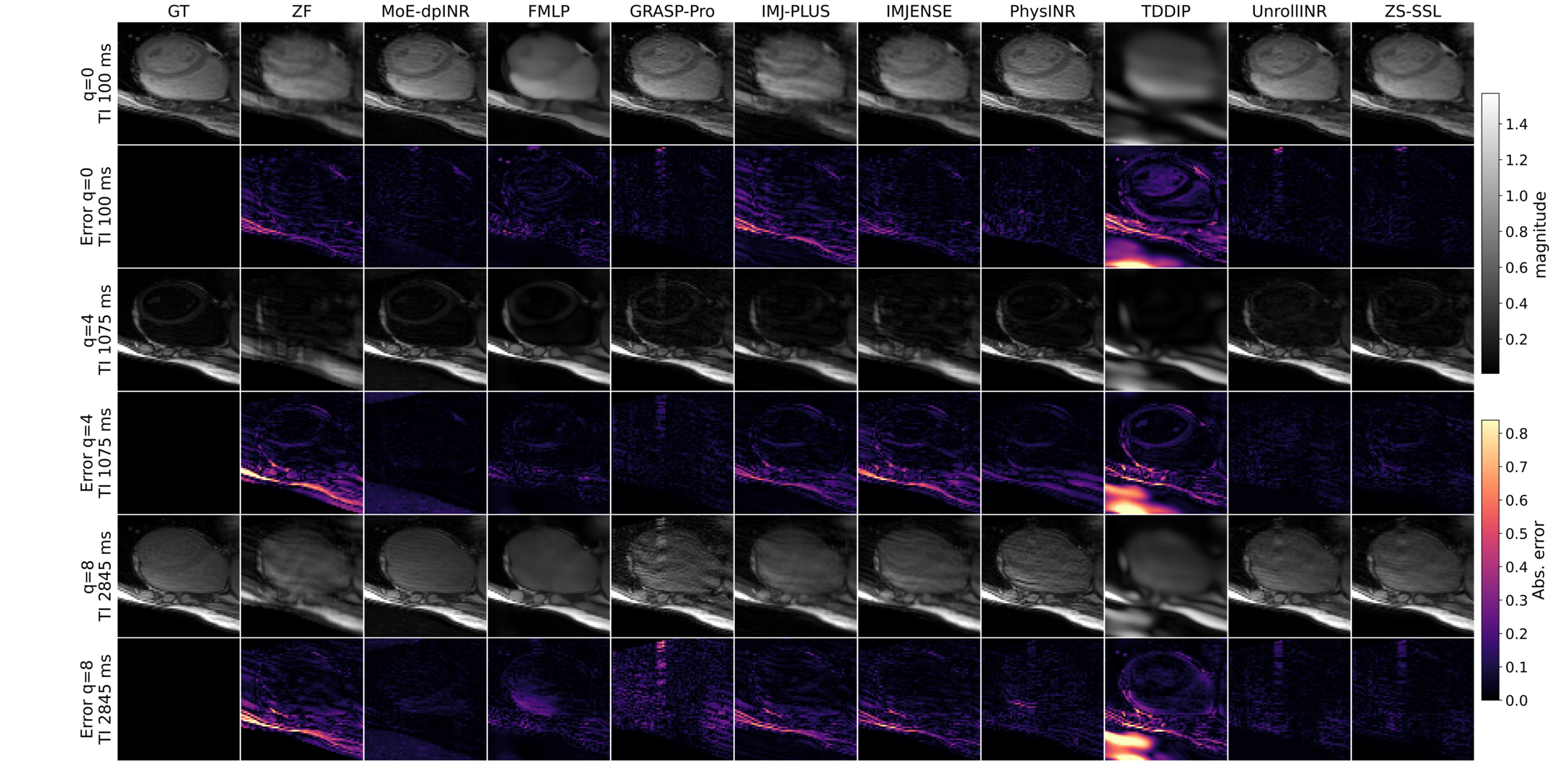} \\[0.5mm]

\rule{0.98\textwidth}{0.4pt} \\[0.5mm]
\textbf{(B) AF=$4\times$: Derived T1 map} \\[0.5mm]
\includegraphics[width=0.98\textwidth]{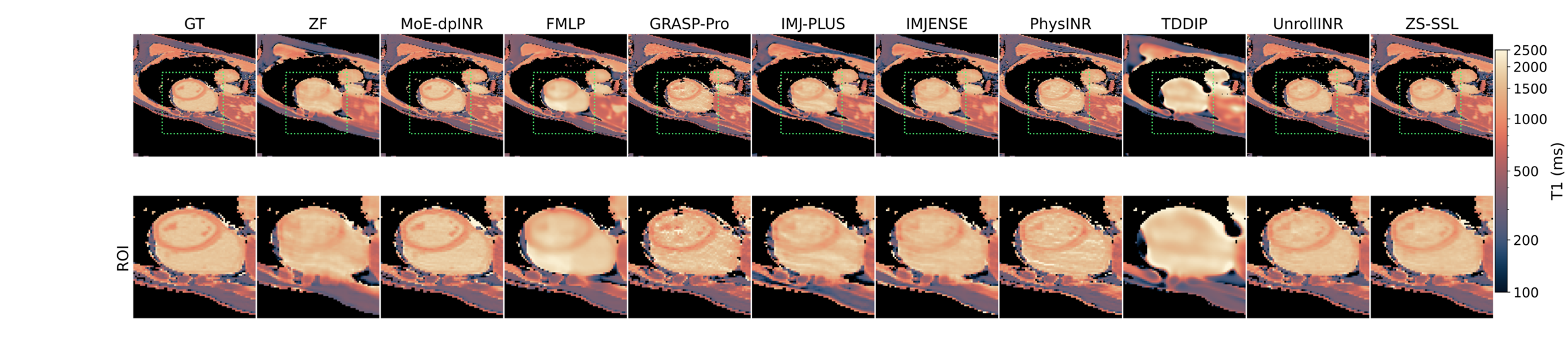} \\[0.5mm]

\rule{0.98\textwidth}{0.4pt} \\[0.5mm]
\textbf{(C) AF=$6\times$: T1-weighted image reconstruction} \\[0.5mm]
\includegraphics[width=0.98\textwidth]{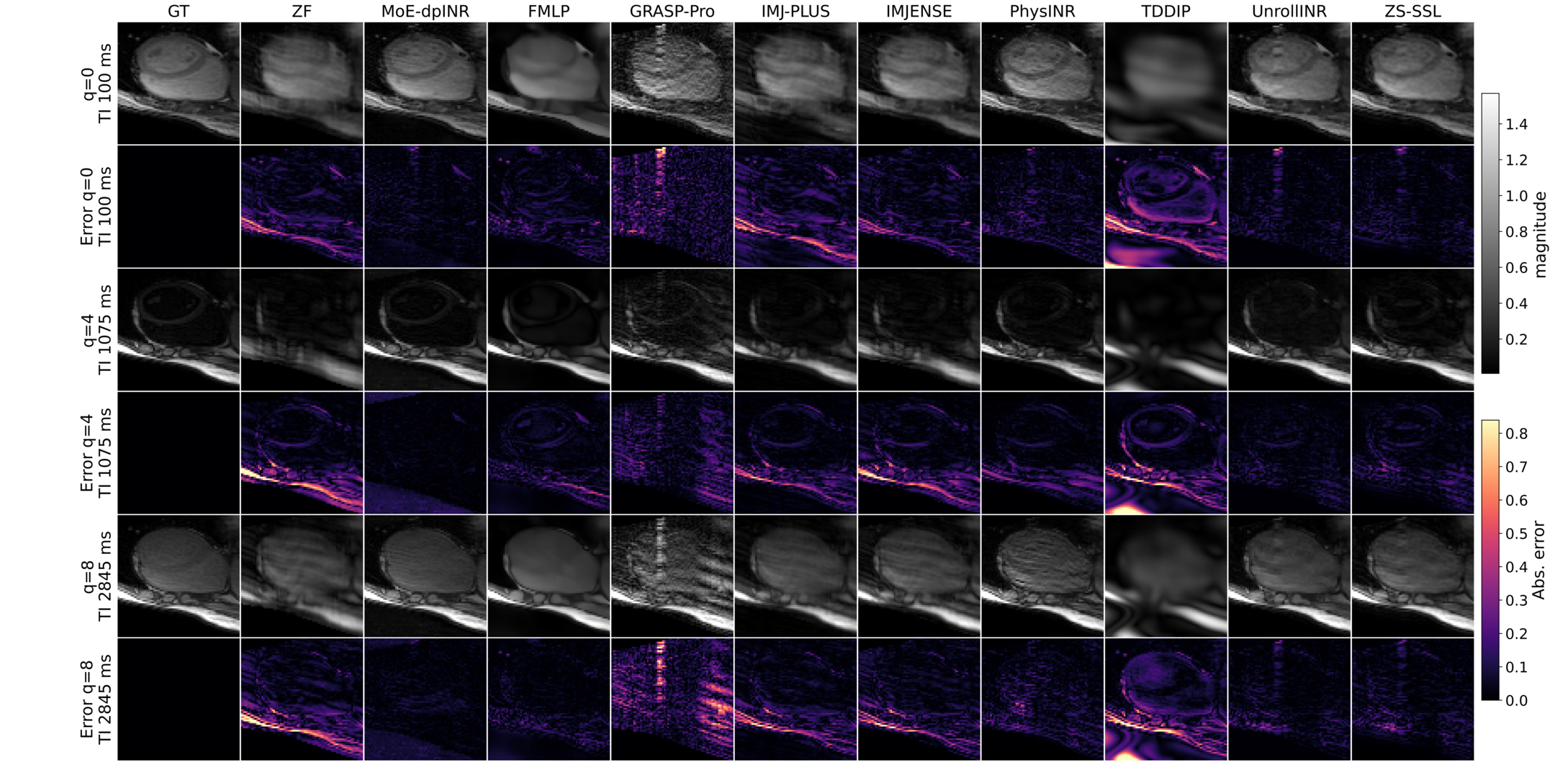} \\[0.5mm]

\rule{0.98\textwidth}{0.4pt} \\[0.5mm]
\textbf{(D) AF=$6\times$: Derived T1 map} \\[0.5mm]
\includegraphics[width=0.98\textwidth]{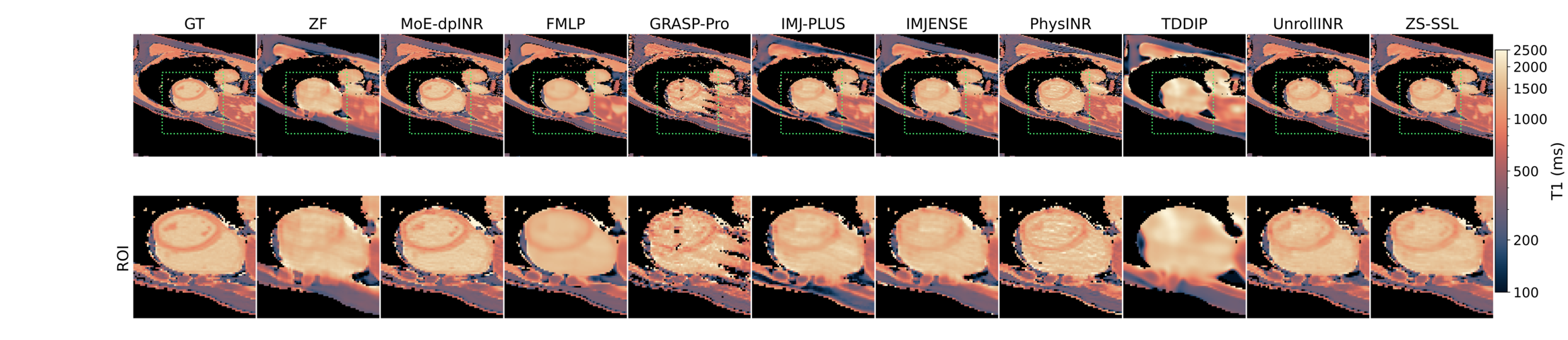}
\end{tabular}
\end{adjustbox}

\caption{
Representative T1-mapping qualitative reconstruction comparison at (A, B) AF=$4\times$ and (C, D) AF=$6\times$. (A, C) T1-weighted image reconstruction (Cardiac ROI) results for representative ordered contrast states. (B, D) Derived T1 maps from the reconstructed T1-weighted image series. MoE-dqINR better preserves contrast-dependent local structure and yields T1 maps closer to the fully sampled reference than the compared baselines.
}
\label{fig:t1m-qualitative-roi-map}
\end{figure*}

Figures~\ref{fig:cine-qualitative-roi} and \ref{fig:t1m-qualitative-roi-map} show representative qualitative reconstruction results at AF=$4\times$ and AF=$6\times$. For cine MRI, the ROI views and $x$--$t$ cross-sections highlight local anatomical fidelity and temporal consistency across the cardiac cycle. For T1 mapping, the ROI T1-weighted images assess contrast-state reconstruction fidelity, while the derived T1 maps evaluate whether the reconstructed image series preserves quantitative parameter estimation. Across both modalities and acceleration factors, MoE-dqINR preserves sharper local structure, reduces localized artifacts, and produces T1 maps closer to the fully sampled reference than the compared baselines, consistent with the quantitative results in Tables~\ref{tab:cine-main-results} and \ref{tab:t1m-main-results}.

\begin{table*}[!t]
\centering
\caption{Cine quantitative comparison across acceleration factors. PSNR $\uparrow$, SSIM $\uparrow$, LPIPS $\downarrow$. Best values are shown in bold, and $^{*}$ indicates statistically significant difference from MoE-dqINR by Wilcoxon test ($p<0.05$).}
\label{tab:cine-main-results}
\resizebox{\textwidth}{!}{
\begin{tabular}{lc|lll|lll|lll}
\hline
Method & Year & \multicolumn{3}{c|}{AF $4\times$} & \multicolumn{3}{c|}{AF $6\times$} & \multicolumn{3}{c}{AF $8\times$} \\
 &  & PSNR $\uparrow$ & SSIM $\uparrow$ & LPIPS $\downarrow$ & PSNR $\uparrow$ & SSIM $\uparrow$ & LPIPS $\downarrow$ & PSNR $\uparrow$ & SSIM $\uparrow$ & LPIPS $\downarrow$ \\
\hline
MoE-dqINR & -- & \textbf{37.31} & \textbf{0.9523} & \textbf{0.0350} & \textbf{35.77} & \textbf{0.9411} & \textbf{0.0428} & \textbf{34.32} & \textbf{0.9228} & \textbf{0.0566} \\
IMJ-PLUS & 2026 & 33.73$^{*}$ & 0.9047$^{*}$ & 0.0886$^{*}$ & 32.07$^{*}$ & 0.8780$^{*}$ & 0.1200$^{*}$ & 30.90$^{*}$ & 0.8595$^{*}$ & 0.1434$^{*}$ \\
MoCo-INR & 2026 & 35.28$^{*}$ & 0.9406$^{*}$ & 0.0410$^{*}$ & 33.80$^{*}$ & 0.9241$^{*}$ & 0.0519 & 32.21$^{*}$ & 0.8996$^{*}$ & 0.0701 \\
ST-INR & 2025 & 34.30$^{*}$ & 0.9230$^{*}$ & 0.0646$^{*}$ & 32.60$^{*}$ & 0.9045$^{*}$ & 0.0802$^{*}$ & 31.07$^{*}$ & 0.8873$^{*}$ & 0.0966$^{*}$ \\
UnrollINR & 2025 & 29.77$^{*}$ & 0.8515$^{*}$ & 0.1182$^{*}$ & 28.09$^{*}$ & 0.8219$^{*}$ & 0.1426$^{*}$ & 26.96$^{*}$ & 0.8030$^{*}$ & 0.1625$^{*}$ \\
FMLP & 2024 & 34.94$^{*}$ & 0.9305$^{*}$ & 0.1037$^{*}$ & 33.50$^{*}$ & 0.9147$^{*}$ & 0.1197$^{*}$ & 32.46$^{*}$ & 0.9017$^{*}$ & 0.1310$^{*}$ \\
IMJENSE & 2024 & 32.40$^{*}$ & 0.8935$^{*}$ & 0.1126$^{*}$ & 30.49$^{*}$ & 0.8606$^{*}$ & 0.1511$^{*}$ & 29.54$^{*}$ & 0.8426$^{*}$ & 0.1752$^{*}$ \\
ZS-SSL & 2022 & 33.74$^{*}$ & 0.9183$^{*}$ & 0.0632$^{*}$ & 31.57$^{*}$ & 0.8885$^{*}$ & 0.0866$^{*}$ & 29.94$^{*}$ & 0.8652$^{*}$ & 0.1088$^{*}$ \\
TDDIP & 2021 & 33.20$^{*}$ & 0.9010$^{*}$ & 0.0899$^{*}$ & 30.96$^{*}$ & 0.8708$^{*}$ & 0.1136$^{*}$ & 29.37$^{*}$ & 0.8521$^{*}$ & 0.1257$^{*}$ \\
GRASP-Pro & 2020 & 34.64$^{*}$ & 0.9234$^{*}$ & 0.0539$^{*}$ & 31.69$^{*}$ & 0.8760$^{*}$ & 0.0909$^{*}$ & 29.72$^{*}$ & 0.8386$^{*}$ & 0.1263$^{*}$ \\
\hline
\end{tabular}
}
\end{table*}

\begin{table*}[!t]
\centering
\caption{T1-mapping quantitative comparison across acceleration factors. Per-state PSNR $\uparrow$, SSIM $\uparrow$, and LPIPS $\downarrow$, and T1-map NMSE $\downarrow$. Best values are shown in bold, and $^{*}$ indicates statistically significant difference from MoE-dqINR by Wilcoxon test ($p<0.05$).}
\label{tab:t1m-main-results}
\resizebox{\textwidth}{!}{
\begin{tabular}{lc|llll|llll|llll}
\hline
Method & Year & \multicolumn{4}{c|}{AF $4\times$} & \multicolumn{4}{c|}{AF $6\times$} & \multicolumn{4}{c}{AF $8\times$} \\
 &  & PSNR $\uparrow$ & SSIM $\uparrow$ & LPIPS $\downarrow$ & Map NMSE $\downarrow$ & PSNR $\uparrow$ & SSIM $\uparrow$ & LPIPS $\downarrow$ & Map NMSE $\downarrow$ & PSNR $\uparrow$ & SSIM $\uparrow$ & LPIPS $\downarrow$ & Map NMSE $\downarrow$ \\
\hline
MoE-dqINR & -- & \textbf{35.32} & \textbf{0.9169} & \textbf{0.0419} & \textbf{0.2590} & \textbf{34.26} & \textbf{0.8933} & \textbf{0.0576} & \textbf{0.2780} & \textbf{32.97} & \textbf{0.8702} & \textbf{0.0736} & \textbf{0.3044} \\
IMJ-PLUS & 2026 & 27.56$^{*}$ & 0.7940$^{*}$ & 0.1240$^{*}$ & 0.4006$^{*}$ & 26.51$^{*}$ & 0.7652$^{*}$ & 0.1574$^{*}$ & 0.4297$^{*}$ & 26.08$^{*}$ & 0.7530$^{*}$ & 0.1767$^{*}$ & 0.4437$^{*}$ \\
PhysINR & 2026 & 31.38$^{*}$ & 0.8839$^{*}$ & 0.0632$^{*}$ & 0.3484$^{*}$ & 29.98$^{*}$ & 0.8556$^{*}$ & 0.0794$^{*}$ & 0.4368$^{*}$ & 28.84$^{*}$ & 0.8253$^{*}$ & 0.0967$^{*}$ & 0.5289$^{*}$ \\
UnrollINR & 2025 & 32.20$^{*}$ & 0.8881$^{*}$ & 0.0726$^{*}$ & 0.3338$^{*}$ & 31.53$^{*}$ & 0.8729$^{*}$ & 0.0845$^{*}$ & 0.3667$^{*}$ & 30.43$^{*}$ & 0.8529$^{*}$ & 0.1020$^{*}$ & 0.3902$^{*}$ \\
FMLP & 2024 & 32.00$^{*}$ & 0.8862$^{*}$ & 0.1131$^{*}$ & 0.3317$^{*}$ & 31.32$^{*}$ & 0.8712$^{*}$ & 0.1255$^{*}$ & 0.3660$^{*}$ & 30.87$^{*}$ & 0.8612 & 0.1324$^{*}$ & 0.3721$^{*}$ \\
IMJENSE & 2024 & 28.81$^{*}$ & 0.8368$^{*}$ & 0.1273$^{*}$ & 0.4592$^{*}$ & 27.80$^{*}$ & 0.8121$^{*}$ & 0.1575$^{*}$ & 0.4897$^{*}$ & 27.16$^{*}$ & 0.7946$^{*}$ & 0.1803$^{*}$ & 0.4897$^{*}$ \\
ZS-SSL & 2022 & 33.15$^{*}$ & 0.9034$^{*}$ & 0.0560$^{*}$ & 0.3129$^{*}$ & 31.82$^{*}$ & 0.8786$^{*}$ & 0.0697$^{*}$ & 0.3745$^{*}$ & 30.66$^{*}$ & 0.8586$^{*}$ & 0.0860$^{*}$ & 0.4072$^{*}$ \\
TDDIP & 2021 & 26.51$^{*}$ & 0.7370$^{*}$ & 0.3053$^{*}$ & 0.5471$^{*}$ & 26.38$^{*}$ & 0.7312$^{*}$ & 0.3096$^{*}$ & 0.5251$^{*}$ & 26.46$^{*}$ & 0.7348$^{*}$ & 0.3090$^{*}$ & 0.5276$^{*}$ \\
GRASP-Pro & 2020 & 30.30$^{*}$ & 0.8248$^{*}$ & 0.1130$^{*}$ & 0.5229$^{*}$ & 27.42$^{*}$ & 0.7519$^{*}$ & 0.1684$^{*}$ & 0.6828$^{*}$ & 25.37$^{*}$ & 0.7053$^{*}$ & 0.2053$^{*}$ & 0.7377$^{*}$ \\
\hline
\end{tabular}
}
\end{table*}

\begin{table*}[!t]
\centering
\caption{Additional low-field OCMR 0.55\,T cine quantitative comparison across acceleration factors. PSNR $\uparrow$, SSIM $\uparrow$, LPIPS $\downarrow$. Best values are shown in bold, and $^{*}$ indicates statistically significant difference from MoE-dqINR by Wilcoxon test ($p<0.05$).}
\label{tab:ocmr-lowfield-results}
\resizebox{\textwidth}{!}{
\begin{tabular}{lc|lll|lll|lll}
\hline
Method & Year & \multicolumn{3}{c|}{AF $2\times$} & \multicolumn{3}{c|}{AF $4\times$} & \multicolumn{3}{c}{AF $6\times$} \\
 &  & PSNR $\uparrow$ & SSIM $\uparrow$ & LPIPS $\downarrow$ & PSNR $\uparrow$ & SSIM $\uparrow$ & LPIPS $\downarrow$ & PSNR $\uparrow$ & SSIM $\uparrow$ & LPIPS $\downarrow$ \\
\hline
MoE-dqINR & -- & \textbf{34.26} & \textbf{0.9197} & \textbf{0.0389} & \textbf{31.86} & \textbf{0.8901} & \textbf{0.0615} & \textbf{30.63} & \textbf{0.8727} & \textbf{0.0752} \\
IMJ-PLUS & 2026 & 32.65$^{*}$ & 0.8996$^{*}$ & 0.0520$^{*}$ & 29.19$^{*}$ & 0.8344$^{*}$ & 0.1010$^{*}$ & 29.01$^{*}$ & 0.8207$^{*}$ & 0.1188$^{*}$ \\
MoCo-INR & 2026 & 28.36$^{*}$ & 0.8145$^{*}$ & 0.0834$^{*}$ & 27.02$^{*}$ & 0.7710$^{*}$ & 0.1105$^{*}$ & 26.26$^{*}$ & 0.7509$^{*}$ & 0.1232$^{*}$ \\
ST-INR & 2025 & 29.82$^{*}$ & 0.8073$^{*}$ & 0.0978$^{*}$ & 26.73$^{*}$ & 0.7222$^{*}$ & 0.1572$^{*}$ & 25.64$^{*}$ & 0.6887$^{*}$ & 0.1839$^{*}$ \\
UnrollINR & 2025 & 30.34$^{*}$ & 0.8203$^{*}$ & 0.0916$^{*}$ & 27.49$^{*}$ & 0.7391$^{*}$ & 0.1573$^{*}$ & 26.69$^{*}$ & 0.7141$^{*}$ & 0.1830$^{*}$ \\
FMLP & 2024 & 31.91$^{*}$ & 0.8455$^{*}$ & 0.2006$^{*}$ & 30.78$^{*}$ & 0.8250$^{*}$ & 0.2227$^{*}$ & 29.98$^{*}$ & 0.8093$^{*}$ & 0.2360$^{*}$ \\
IMJENSE & 2024 & 28.53$^{*}$ & 0.8659$^{*}$ & 0.0852$^{*}$ & 26.92$^{*}$ & 0.8007$^{*}$ & 0.1593$^{*}$ & 26.25$^{*}$ & 0.7745$^{*}$ & 0.1966$^{*}$ \\
ZS-SSL & 2022 & 31.33$^{*}$ & 0.8786$^{*}$ & 0.0703$^{*}$ & 29.09$^{*}$ & 0.8143$^{*}$ & 0.1303$^{*}$ & 27.97$^{*}$ & 0.7772$^{*}$ & 0.1709$^{*}$ \\
TDDIP & 2021 & 27.37$^{*}$ & 0.7306$^{*}$ & 0.1598$^{*}$ & 24.44$^{*}$ & 0.6428$^{*}$ & 0.2542$^{*}$ & 22.03$^{*}$ & 0.5719$^{*}$ & 0.2894$^{*}$ \\
GRASP-Pro & 2020 & 31.71$^{*}$ & 0.8635$^{*}$ & 0.0504$^{*}$ & 28.37$^{*}$ & 0.7595$^{*}$ & 0.1158$^{*}$ & 26.52$^{*}$ & 0.7000$^{*}$ & 0.1658$^{*}$ \\
\hline
\end{tabular}
}
\end{table*}

\subsection{State-Conditioned Routing Analysis}\label{subsec:routing-analysis}

\begin{figure*}[!t]
\centering
\newlength{\routinganimheight}
\setlength{\routinganimheight}{0.33\textwidth}
\begin{tabular}{@{}c|c@{}}
\begin{minipage}[t]{0.44\textwidth}
\centering
\textbf{(A) Cine}\par\medskip
\includegraphics[height=\routinganimheight]{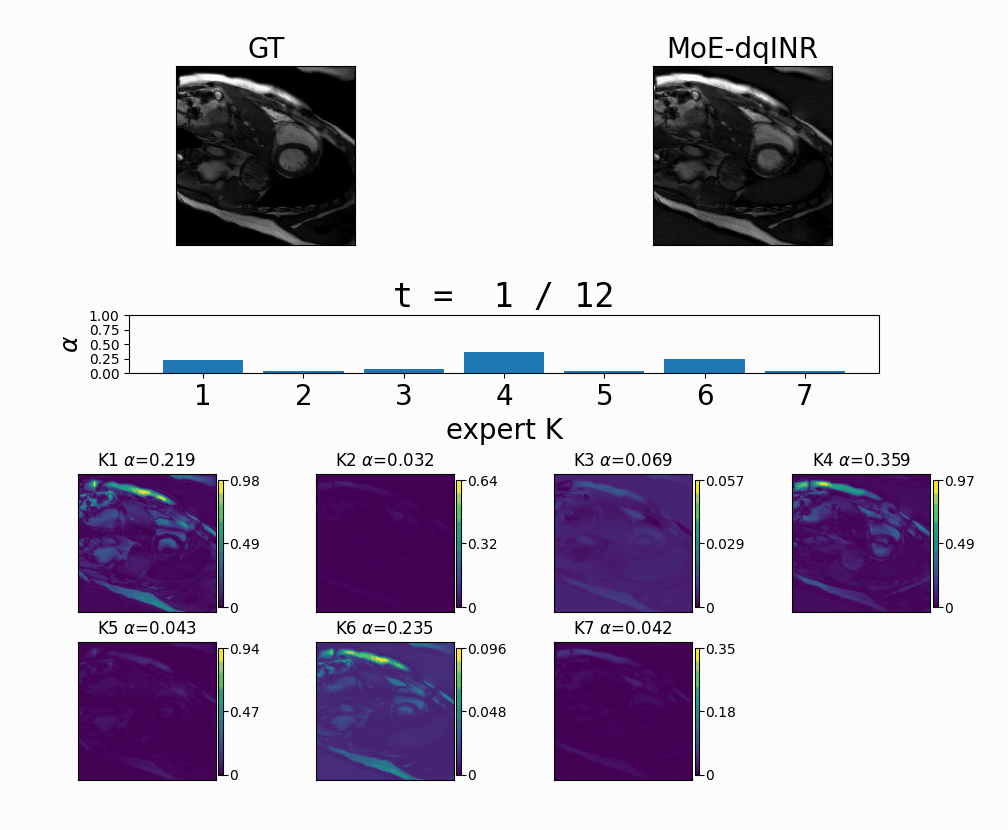}\par\medskip
\includegraphics[width=\linewidth]{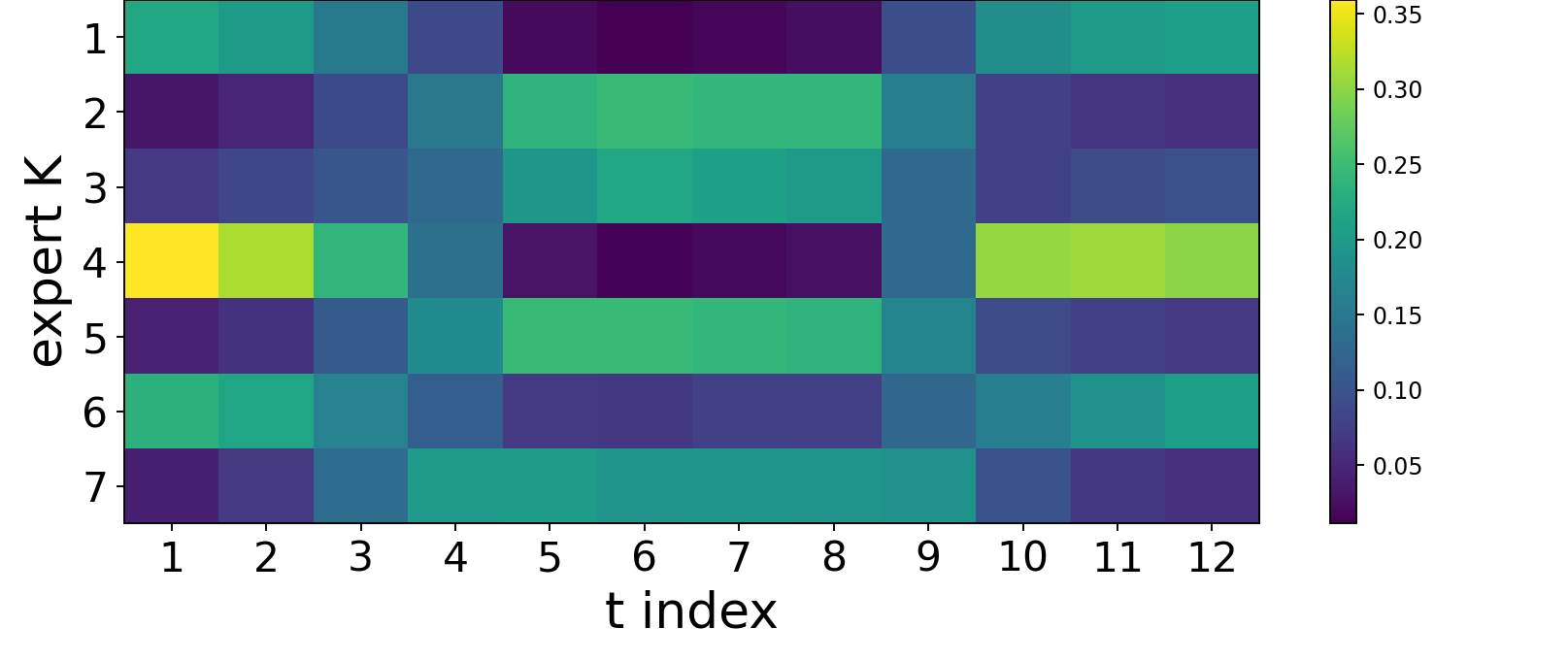}
\end{minipage}
&
\begin{minipage}[t]{0.44\textwidth}
\centering
\textbf{(B) T1 Mapping}\par\medskip
\includegraphics[height=\routinganimheight]{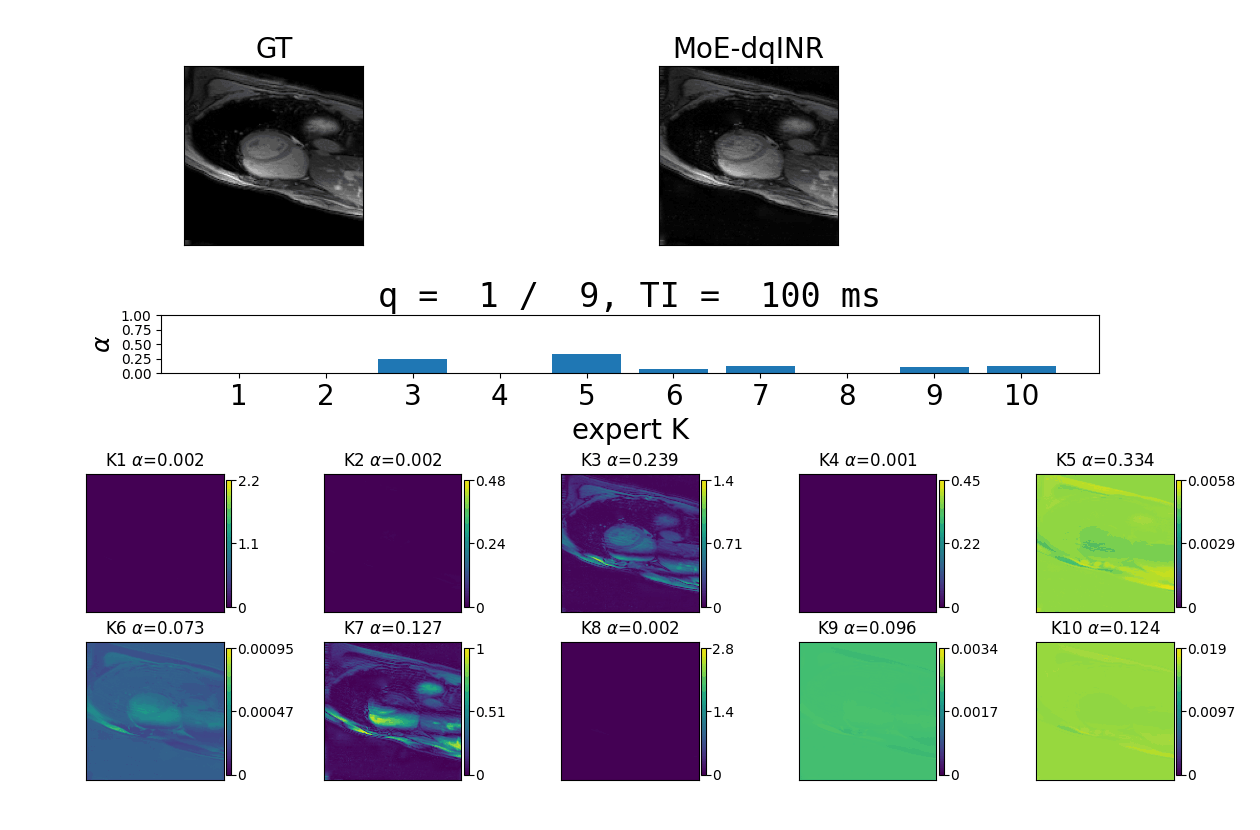}\par\medskip
\includegraphics[width=\linewidth]{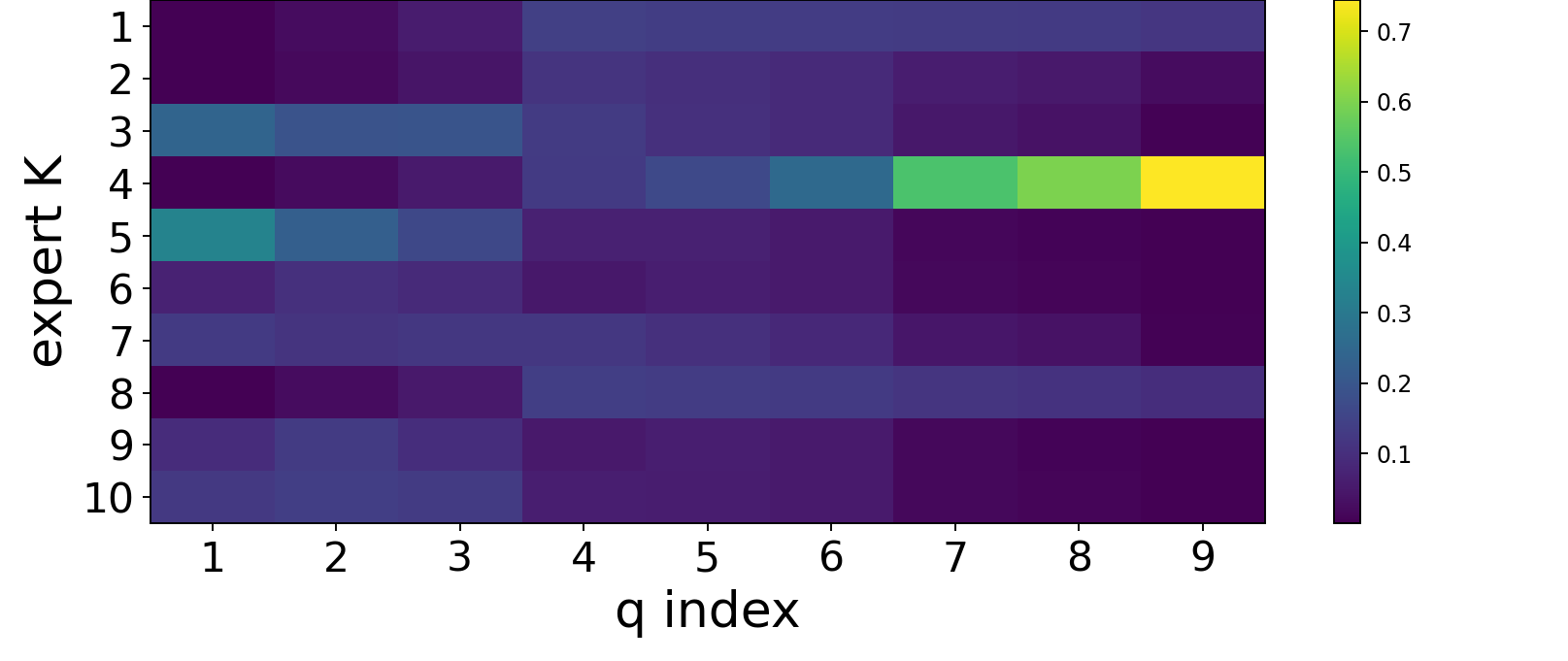}
\end{minipage}
\end{tabular} \\[5mm]
\caption{
\textbf{Top:} State-conditioned expert routing behavior in representative cine and T1-mapping reconstructions at AF=$6\times$. (A) For cine, the router combines $K=7$ shared spatial experts across 12 temporal frames; (B) for T1 mapping, it combines $K=10$ shared spatial experts across 9 ordered contrast states. \textbf{Bottom:} The corresponding routing matrices show the simplex-normalized expert weights $\alpha_{qk}$, with rows indexing experts and columns indexing the ordered state axis: (A) temporal frame $t$ for cine and (B) contrast-state index $q$, ordered by inversion time, for T1 mapping. (A) In cine, the routing weights remain distributed across multiple experts and exhibit an approximately cyclic pattern across temporal frames, indicating reuse of shared spatial experts across temporally related cardiac phases. (B) In T1 mapping, the routing is more selective: a small subset of experts carries most of the probability mass, and the dominant expert changes as the ordered contrast state increases. This suggests that the same state-conditioned routing mechanism adapts to distributed periodic mixing in dynamic MRI and contrast-selective expert handover in qMRI.
}
\label{fig:expert-routing-analysis}
\vspace{10mm}
\end{figure*}

Figure~\ref{fig:expert-routing-analysis} visualizes the learned routing matrices for representative cine and T1-mapping reconstructions at AF=$6\times$. This analysis is placed in the Results section because the routing matrices are learned outcomes of scan-specific optimization rather than components required to define the method.

In the cine case, the simplex-normalized weights are distributed across several of the $K=7$ experts, and no single expert dominates the synthesis across the cardiac cycle. The routing pattern is also mostly symmetric about the central temporal axis and approximately cyclic: early and late frames activate similar expert combinations, while intermediate frames shift toward different mixtures. This behavior is consistent with ECG-gated cine dynamics, in which the cardiac cycle progresses from end-systole to end-diastole and back toward end-systole, causing anatomically related cardiac phases to recur over time. The result suggests that MoE-dqINR reuses shared spatial experts both within each frame and across temporally related frames, rather than assigning each frame to an independent expert.

In contrast, the T1-mapping case shows a more selective routing pattern over the $K=10$ experts. Because the contrast states are ordered by inversion time, increasing $q$ corresponds to progression along the T1-weighted contrast evolution, although the router itself receives the normalized state index rather than the raw TI value. Most routing mass is concentrated in a small number of experts, and the leading expert changes systematically as $q$ increases. At smaller $q$, corresponding to lower TI values, the routing distribution is dominated by a few experts. As the longitudinal magnetization evolves toward darker contrast and approaches the null point as $q$ grows, the weights assigned to these initially dominant experts gradually diminish. After the null point, where the T1-weighted images begin to brighten again as longitudinal magnetization recovers, dominance is transferred to a different set of experts, where the earlier dominant experts decay toward zero, while the newly dominant experts increase in weight with increasing $q$. This handover indicates that the router allocates different shared spatial experts to different regimes of the contrast-evolution trajectory. The remaining low-weight experts may provide lower-amplitude corrective components for local detail, boundaries, or contrast-specific residual structure, rather than carrying the dominant contrast change.

\subsection{Dynamic Temporal Residual Branch Ablation}\label{subsec:dynamic-temporal-residual-branch-ablation}

\begin{table}[!t]
\centering
\caption{Dynamic temporal residual-branch ablation for cine reconstruction across acceleration factors. PSNR $\uparrow$, SSIM $\uparrow$, LPIPS $\downarrow$. Means are reported. Best values are shown in bold, and $^{*}$ indicates statistically significant difference from MoE-dqINR by Wilcoxon test ($p<0.05$).}
\label{tab:cine-residual-ablation}
\small
\setlength{\tabcolsep}{4.2pt}
\begin{tabular}{lllll}
\hline
AF & Variant & PSNR $\uparrow$ & SSIM $\uparrow$ & LPIPS $\downarrow$ \\
\hline
\multirow{3}{*}{$4\times$}
 & MoE + Residual & \textbf{37.31} & \textbf{0.9523} & \textbf{0.0350} \\
 & Residual only & 35.22$^{*}$ & 0.9317$^{*}$ & 0.0575$^{*}$ \\
 & MoE w/o Residual & 36.89 & 0.9458 & 0.0403 \\
\hline
\multirow{3}{*}{$6\times$}
 & MoE + Residual & \textbf{35.77} & \textbf{0.9411} & \textbf{0.0428} \\
 & Residual only & 33.95$^{*}$ & 0.9138$^{*}$ & 0.0687$^{*}$ \\
 & MoE w/o Residual & 35.53$^{*}$ & 0.9379$^{*}$ & 0.0454 \\
\hline
\multirow{3}{*}{$8\times$}
 & MoE + Residual & \textbf{34.32} & \textbf{0.9228} & \textbf{0.0566} \\
 & Residual only & 32.40$^{*}$ & 0.8853$^{*}$ & 0.0885$^{*}$ \\
 & MoE w/o Residual & 34.18$^{*}$ & 0.9189 & 0.0588 \\
\hline
\end{tabular}
\end{table}

We further evaluated the effect of the dynamic temporal residual branch in cine reconstruction. Three variants were compared in Table~\ref{tab:cine-residual-ablation}: the full MoE-dqINR model with the residual branch enabled ("MoE + Residual"), a residual-only variant in which the compact residual pathway replaces the main expert-mixture pathway ("Residual only"), and a no-residual variant that keeps the MoE backbone but removes the residual branch ("MoE w/o Residual"). The full "MoE + Residual" configuration performs best across all cine metrics, confirming that the residual path is most effective when it complements, rather than replaces, the shared-expert representation. The residual-only variant ("Residual only") performs worst, indicating that a low-capacity residual pathway alone is insufficient to model the full spatial-temporal variability.

\subsection{Timing Comparison}\label{subsec:timing-comparison}

We also compared per-scan optimization and inference time across method families. Tables~\ref{tab:cine-timing} and \ref{tab:t1m-timing} group the baselines into a conventional (compressed sensing (CS)-based) method, self-supervised non-INR learning methods, and self-supervised INR-based methods.

As expected for a conventional model-based method without neural-field fitting, GRASP-Pro was the fastest baseline in absolute optimisation time. However, this speed came with substantially poorer reconstruction performance compared to all other baselines (Table~\ref{tab:cine-main-results} and Table~\ref{tab:t1m-main-results}).

However, across all evaluated scan-specific learned baselines, MoE-dqINR achieves the lowest per-scan optimization time across both tasks, requiring 31.08 s for cine MRI and 26.82 s for T1-mapping, compared with 102.25--1446.98 s for the other INR and self-supervised deep learning baselines reported in Tables~\ref{tab:cine-timing} and \ref{tab:t1m-timing}.

Within the learned scan-specific baselines, MoE-dqINR provided the strongest efficiency--performance trade-off, combining the best reported reconstruction metrics with per-scan optimisation times of around 30 seconds. The proposed method therefore combines substantially stronger reconstruction performance with markedly lower scan-specific optimization cost than the evaluated INR alternatives, while also maintaining low inference time relative to the broader family of learned baselines.

\begin{table}[!t]
\centering
\caption{Cine per-scan optimization and inference time (Unit: seconds)}
\label{tab:cine-timing}
\small
\begin{tabular}{p{2.1cm}p{2.0cm}r@{\hspace{0.9em}}r}
\hline
Category & Method & Optimization & Inference \\
\hline
\multirow{1}{*}{\shortstack[l]{Conventional}} & GRASP-Pro & 0.56 & 0.02 \\
\hline
\multirow{2}{*}{\shortstack[l]{Self-supervised\\(non-INR)}} & ZS-SSL & 1446.98 & 1.81 \\
 & TDDIP & 250.68 & <0.01 \\
\hline
\multirow{7}{*}{\shortstack[l]{Self-supervised\\(INR)}} & MoE-dqINR & \textbf{31.08} & \textbf{<0.01} \\
 & IMJ-PLUS & 164.29 & \textbf{<0.01} \\
 & MoCo-INR & 205.00 & \textbf{<0.01} \\
 & ST-INR & 263.82 & \textbf{<0.01} \\
 & UnrollINR & 1059.43 & 0.14 \\
 & FMLP & 397.27 & 0.21 \\
 & IMJENSE & 471.56 & 0.21 \\
\hline
\end{tabular}
\end{table}

\begin{table}[!t]
\centering
\caption{T1-mapping per-scan optimization and inference time (Unit: seconds)}
\label{tab:t1m-timing}
\small
\begin{tabular}{p{2.1cm}p{2.0cm}r@{\hspace{0.9em}}r}
\hline
Category & Method & Optimization & Inference \\
\hline
\multirow{1}{*}{\shortstack[l]{Conventional}} & GRASP-Pro & 1.78 & <0.01 \\
\hline
\multirow{2}{*}{\shortstack[l]{Self-supervised\\(non-INR)}} & ZS-SSL & 1274.12 & 0.72 \\
 & TDDIP & 124.60 & <0.01 \\
\hline
\multirow{6}{*}{\shortstack[l]{Self-supervised\\(INR)}} & MoE-dqINR & \textbf{26.82} & \textbf{<0.01} \\
 & IMJ-PLUS & 102.25 & \textbf{<0.01} \\
 & PhysINR & 706.91 & 0.14 \\
 & UnrollINR & 507.57 & 0.06 \\
 & FMLP & 364.79 & 0.09 \\
 & IMJENSE & 167.71 & 0.07 \\
\hline
\end{tabular}
\end{table}

\section{Discussion}\label{sec:discussion}

The results support the central design hypothesis of this work: separating reusable spatial representation from state-dependent synthesis provides a strong scan-specific prior for both cine MRI and T1-mapping reconstruction. Across the main quantitative comparisons, MoE-dqINR consistently outperforms the compared baselines for both tasks, indicating that the proposed shared-expert image representation and state-conditioned routing capture cross-state common structure without collapsing state-specific variation into a single monolithic coordinate field. This is particularly relevant for the present two-task setting, where the non-spatial axis has different semantics in the two modalities. In cine, the ordered state axis corresponds to temporal frames; in T1-mapping, it corresponds to ordered contrast states.

The expert-routing analysis in Fig.~\ref{fig:expert-routing-analysis} provides qualitative insight into this behavior. In cine reconstruction, routing weights are broadly distributed and approximately cyclic across cardiac phases, suggesting that the model reuses spatial experts across related motion states and reflects the periodic structure of cardiac dynamics. T1 mapping instead shows more concentrated routing, with a smaller subset of experts take turns to become dominant at different inversion-time states. This difference suggests that the router adapts its expert combinations to the semantics of the state axis: distributed expert reuse emerges for cyclic cardiac motion, whereas more selective expert use emerges for ordered contrast evolution.

Notably, these routing matrices should be interpreted as qualitative evidence of learned functional specialization rather than as anatomically labeled decompositions. The experts are optimized only through multicoil k-space data consistency, and their roles are not preassigned. They may therefore capture shared anatomy, contrast-dependent structure, boundary corrections, local residual detail, or combinations of these factors. Nevertheless, the contrast between distributed cine routing and selective T1-mapping routing supports the intended role of the router. Rather than relying on modality-specific inductive bias in the form of a fixed temporal or quantitative signal model, the router learns how to combine reusable spatial experts according to the acquisition-state structure. Together with the effectiveness of the same factorized formulation across both tasks, these observations support the main design claim that separating shared spatial representation from state-dependent synthesis provides a unified but modality-adaptive scan-specific reconstruction prior.

The additional 0.55\,T OCMR validation further strengthens this interpretation: even under a lower-field cine acquisition setting that differs from the main benchmark, the proposed method remains best across all tested acceleration factors and all three image-quality metrics. This indicates that the learned shared-expert decomposition is not narrowly tied to a single scanner regime or field strength.

The ablation results clarify the contribution of the residual branch. In cine reconstruction, the best performance is obtained when the residual branch is added on top of the MoE backbone, whereas the residual-only and no-residual variants are both inferior. This indicates that the residual path is not acting as a replacement for the expertized image representation, but as a low-capacity corrective pathway that refines frame-specific deviations around a stronger shared-expert model.

The timing results position the method from a practical perspective. Conventional CS-based reconstruction, represented here by GRASP-Pro, remains faster in absolute per-scan optimization time, which is expected given its different optimization structure and the absence of learned neural-field components. However, within the learned baselines, MoE-dqINR offers a favorable efficiency--performance tradeoff. Across both tasks, it achieves the lowest per-scan optimization time among the evaluated scan-specific INR baselines, requiring 31.08 s for cine MRI and 26.82 s for T1-mapping, compared with 102.25--1059.43 s for the other INR baselines. This combination matters for scan-specific methods, where per-scan optimization cost is part of the deployment burden. The results therefore suggest that the proposed factorization improves not only reconstruction accuracy but also optimization efficiency relative to more entangled INR formulations.

Several limitations should be noted. First, although the multi-center and multi-vendor experiment setting broadens the evaluation, the study still focuses on retrospective undersampling and on short-axis cine and T1-mapping, so broader validation across other sequences, and acquisition settings may be recommended \cite{wang2025cmrxrecon2024}. Second, although the model is unified across cine and T1-mapping, it remains scan-specific and requires per-scan optimization. Third, the present formulation reconstructs image series directly rather than jointly estimating calibration variables, deformation fields, or quantitative parameter maps inside a single end-to-end framework. These choices keep the model focused and interpretable, but they also leave room for future extensions. Natural next steps include prospective validation, extension to additional qMRI contrasts, integration with calibration-aware or motion-aware components when needed, and investigation of whether the same shared-expert routing principle can be combined with parameter-map-consistent quantitative modeling without losing the image-first advantages demonstrated here.

\bibliographystyle{unsrtnat}
\bibliography{MoE-INR}

\end{document}